\documentclass[12pt,preprint]{aastex} 

\usepackage{subfigure}
\usepackage{ulem}
\usepackage{color}
\usepackage{graphics,graphicx}
\usepackage{times} 
\usepackage{amssymb}
\usepackage{amsmath}
\usepackage{lscape}

\usepackage{multirow}
\usepackage{ctable}
\usepackage{threeparttable}
\newif\ifAMStwofonts
\AMStwofontstrue
\def\aap{A\&A}                   % "Astron. Astrophys."
\def\apjs{ApJS}                  % "Astrophys. J. Suppl. Ser."
\def\apjl{ApJ}                   % letter at ApJ

\def\rg{${\it r}_{\rm g}$}
\def\rin{${\it r}_{\rm in}$}
\def\rbr{${\it r}_{\rm br}$}

\def\nh{${\it N}_{\rm H}$}
\def\ka{$\rm K\alpha$}

\def\xmm{{\it XMM-Newton~\/}}

\def\suzaku{{\it SUZAKU}}

\def\chandra{{\it Chandra}}

\def\epicmos1{{\it EPIC}{\rm-MOS1~\/}}
\def\epicmos2{{\it EPIC}{\rm-MOS2 ~\/}}
\def\epicmos{{\it EPIC}{\rm-MOS}}
\def\xis{{\rm XIS}}
\def\pin{{\rm PIN}}
\def\hxd{{\rm HXD}}

\def\chandra{{\it Chandra}}
\def\xmm{{\it XMM-Newton}}

\def\suzaku{{\it Suzaku}}

\def\heasoftv{\hbox{\rm HEASOFT\thinspace v6.10}}
\def\xselect{\hbox{\rm XSELECT}}
\def\ftool{\hbox{\rm FTOOL}}

\def\s{\hbox{$\rm\thinspace s$}}
\def\ks{\hbox{$\rm\thinspace ks$}}

\def\cm{\hbox{$\rm\thinspace cm$}}

\def\kmps{\hbox{$\rm\thinspace km~s^{-1}$}}
\def\pcmsq{\hbox{$\rm\thinspace cm^{-2}$}}

\def\ev{\hbox{$\rm\thinspace eV$}}
\def\kev{\hbox{$\rm\thinspace keV$}}

\def\ctsps{\hbox{$\rm\thinspace count~s^{-1}$}}

\def\ergpcmsqps{\hbox{$\rm\thinspace erg~cm^{-2}~s^{-1}$}}
\def\ergcmps{\hbox{$\rm\thinspace erg~cm^{-1}~s^{-1}$}}

\def\ergcmps{\hbox{$\rm\thinspace erg~cm~s^{-1}$}}

\def\msun{\hbox{$\rm\thinspace M_{\odot}$}}

\def\chisq{{\chi^{2}}}

\def\reflionx{\rm{\small REFLIONX}}

\def\kdblur{\rm{\small KDBLUR}}

\def\pexrav{\rm{\small PEXRAV}}
\def\gaussian{\rm{\small GAUSSIAN}}
\def\pexmon{\rm{\small PEXMON}}
\def\relconv{\rm{\small RELCONV}}

\def\xselect{\hbox{\rm{\small XSELECT~\/}}}

\def\ftool{\hbox{\rm{\small FTOOL}}}

\def\grppha{\hbox{\rm{\small GRPPHA~\/}}}
\def\rbnpha{\hbox{\rm{\small RBNPHA~\/}}}
\def\rbnrmf{\hbox{\rm{\small RBNRMF~\/}}}
\def\mathpha{\hbox{\rm{\small MATHPHA}}}
\def\addascaspec{\hbox{\rm{\small ADDASCASPEC~\/}}}

\def\hxddtcor{\hbox{\rm{\small HXDDTCOR}}}

\def\mgtime{\hbox{\rm{\small MGTIME}}}
\def\addascaspec{\hbox{\rm{\small ADDASCASPEC}}}

\def\xstar{\hbox{\rm{\small XSTAR~\/}}}
\def\grid25{\hbox{\rm{\small GRID25}}}

\def\ctps{\hbox{$\rm\thinspace count~s^{-1}$}}

\def\pxl{{\rm ~pixels}}

\def\n{\hbox{\rm NGC~3783}}
\def\m{\hbox{\rm MCG--6-30-15}}
\def\h{\hbox{\rm 1H0707-495}}

\def\spin{{$>0.88$}} % Change spin value here

\begin{document}

\title{X-ray spectral variability in NGC~3783}
\author{
R.~C.~Reis\altaffilmark{1,2},
A.~C.~Fabian\altaffilmark{2},
C.~S.~Reynolds\altaffilmark{3,4},
L.~W.~Brenneman\altaffilmark{5}, 
D.~J.~Walton\altaffilmark{2},
M.~Trippe\altaffilmark{3,4},
J.~M.~Miller\altaffilmark{1},
R.~F.~Mushotzky\altaffilmark{3,4},
M.~A.~Nowak\altaffilmark{6}
}
\altaffiltext{1}{Dept. of Astronomy, University of Michigan, Ann Arbor, Michigan~48109~USA}
\altaffiltext{2}{Institute of Astronomy, University of Cambridge, Madingley Rd., Cambridge CB3 0HA, UK}
\altaffiltext{3}{Dept. of Astronomy, University of Maryland, College
  Park, MD~20742~USA}
\altaffiltext{4}{Joint Space Science Institute (JSI), University of Maryland, College
  Park, MD~20742~USA}
\altaffiltext{5}{Harvard-Smithsonian CfA, 60 Garden St. MS-67, Cambridge, MA~02138~USA}
\altaffiltext{6}{MIT Kavli Institute for Astrophysics, Cambridge, MA~02139~USA}

\begin{abstract}
NGC~3783 was observed for approximately 210\ks\ by \suzaku\ and in this time showed significant spectral and flux variability at both short (20\ks) and long (100\ks) time scales. The full observation is found to consist of approximately six ``spectral periods'' where the behaviour of the soft (0.3--1.0\kev) and hard (2--10\kev) bands are somewhat distinct. Using a variety of methods we find that the strong warm absorber present in this source does not change on these time scales, confirming that the broad-band variability is intrinsic to the central source. The time resolved difference-spectra are well modelled with an absorbed powerlaw below 10\kev, but show an additional hard excess at $\approx 20$\kev\ in the latter stages of the observation. This suggests that, in addition to the variable powerlaw, there is a further variable component that varies with time but not monotonically with flux. We show that a likely interpretation is that this further component is associated with variations in the reflection fraction or possibly ionization state of the accretion disk a few gravitational radii from the black hole.
\end{abstract}


\section{Introduction}

Active Galactic Nuclei (AGN) are commonly found to display persistent X-ray spectral and flux variability in a variety of time-scales (e.g. \citealt{McHardy1989}, \citealt{Greenetal1993}). Studies of this broadband variability usually suggest that at least part is due to strong gravitational effects close to the central black hole. Time-resolved spectral analyses of such sources are thus an important tool in the study of the inner regions of the accretion flow around supermassive black holes, providing information on both the geometry and physical evolution of the inner accretion disk and on intrinsic physical parameters such as black hole mass and spin.

It has been shown for a number of Seyfert galaxies, most notably \m\ (see e.g. \citealt{Fabianvaughan03}), that the spectrum above a few \kev\ becomes harder at lower X-ray fluxes. At all flux levels, however, the X-ray spectrum of such systems usually consists of a primary continuum which is accurately approximated as a powerlaw, together with a broad Fe-\ka\ line and cold reflection features (\citealt{Nandra97}). In some cases it is also accompanied by highly ionized warm absorbers and a ``soft excess'' \citep{Halpern1984, Reynolds1997R}.  The most common interpretation for the spectra of such AGN is that the primary X-ray emission irradiates the underlying optically thick material in the innermost part of the accretion disk, resulting in ``reflection signatures" consisting of fluorescent and recombination emission lines as well as absorption features. The most prominent signature is the broad, skewed Fe-\ka\ line (see e.g. \citealt{miller07review} for a recent review of relativistically broadened lines). Furthermore, the soft excess is a natural consequence of reflection from the innermost regions around a black hole for a wide range of ionisation states. 

%Although  different interpretations of AGN spectra exist (see e.g. \citealt{Turner_miller09} ), they have failed to consistently explain their spectral variability (\citealt{Reynolds09}; \citealt{FabZog09}; \citealt{Zoghbi10} and references therein).

Recent spectral variability studies of Seyfert galaxies such as \m\ (\citealt{Miniutti07}), \h\ (\citealt{FabZog09, Zoghbi10}) and NGC~4051 (\citealt{Uttley04}; \citealt{Terashima08}) have found different explanations for the origin of the variability - sometimes in the same object.  Amongst the possibilities are ({\rm i}) the spectrum consists of a soft component (powerlaw) with constant spectral slope and variable flux, together with a constant hard-component associated with strong reflection from the innermost accretion disk (e.g \m; \citealt{Vaughan_Fabian03, Taylor03, Miniutti07}); ({\rm ii}) the soft component consists of a pivoting powerlaw with the photon index increasing with the logarithm of the flux, together with a constant reflection component (e.g. NGC~4051; \citealt{Uttley04}); and ({\rm iii}) a pivoting  powerlaw together with a constant, partially-covered, neutral (cold) reflection (e.g. NGC~4051; \citealt{Terashima08}). \suzaku\ provides an ideal opportunity to discern between these various distinct interpretations due to its broad-band spectral coverage and good signal-to-noise in the full 0.5--50\kev\ range as well as large effective area below $\sim10$\kev.

NGC~3783 is a bright, nearby Seyfert~1 galaxy exhibiting prominent broad emission lines and strong X-ray absorption features. Based on a total of approximately 900\ks\ of \chandra\  observation, various authors concluded that this source contains three zones of ionized absorption covering a large range of ionization (\citealt{Kaspi02, Netzer03, Krongold2003, Krongold2005}). Analyses of the grating data showed little, if any, evidence for changes in the optical depths of the absorbers.  Similar conclusions have also been made using \xmm\ RGS data (\citealt{Blustin02}). We observed \n\ for approximately 340\ks\ with \suzaku\ as part of the {\it Suzaku AGN Spin Survey} Key Project. A detailed analysis of the time averaged spectrum is presented in \citet{Brenneman20113783} (hereafter Paper~1). For the first time in \n, the dimensionless spin parameter for the central black hole was estimated to be \spin\ at the 99~per~cent level of confidence. This was based on the self-consistent modelling of the full reflection features, whilst simultaneously accounting for absorption due to the warm absorber in the line of sight. In this paper we present an examination of the X-ray spectrum and spectral variability of \n\ using this same long \suzaku\ observation with the goal of better understanding the accretion disk flow geometry and evolution of this system. By usage of a variety of techniques we show that the spectral and flux variation seen in \n\ are likely due to a combination of a varying powerlaw (with constant index) as well as a hard-reflection component. Furthermore, we find that the ionization and reflection fraction in the innermost region of the accretion disk are also likely to be varying.

We start in \S{\thinspace\ref{observation}} with specifics of the observation and data reduction procedure. This is then followed by a detailed analysis of the spectral variability (\S{\thinspace\ref{analyses}}) using a combination of flux-flux relations (\S{\thinspace\ref{flux_flux}}), flux-resolved (\S{\thinspace\ref{flux_resolved}}) and time-resolved difference-spectra (\S{\thinspace\ref{time_resolved}}). We apply the results of the previous sections to each individual time period in (\S{\thinspace\ref{individual}}). Our results are then summarised in \S{\thinspace\ref{discussion}}.

\section{Data reduction}
\label{observation}

We observed NGC~3783 for a total of $\sim$340\ks\ (yielding 210\ks\ of good ``on source" exposure) with \suzaku\ (\citealt{SUZAKU})  as part of the {\it Suzaku AGN Spin Survey} Key Project. The observations were performed during the period 10--15 July 2009. The three operating detectors constituting the X-ray Imaging Spectrometer (XIS; \citealt{SUZ_XIS}) were  operated in the ``normal'' clock and the 3x3 and 5x5 editing modes. The observation resulted in a total (co-added) exposure of approximately 420 and 210\ks\ for the front-illuminated (FI) and back-illuminated (BI) instruments, respectively.

Using the latest \heasoftv\ software package we processed the unfiltered event files for each of the \xis\ CCDs and editing modes operational, following the \suzaku\ Data Reduction Guide\footnote{http://heasarc.gsfc.nasa.gov/docs/suzaku/analysis/}. We started by re-running the \suzaku\ pipeline with the latest calibration (2011-02-10), as well as the associated screening criteria files in order to create new cleaned event files. \xselect was used to extract spectral products from these event files. The good time intervals provided by the \xis\ team were employed in all cases to exclude telemetry saturations.  Source  events were extracted from a circular region of 200\pxl\ ($\sim208$\arcsec) radius centred on the point source, and background spectra from another region of the same size, devoid of any obvious contaminating emission. The script {\hbox{\rm{\small XISRESP}}}{\footnote  {http://suzaku.gsfc.nasa.gov/docs/suzaku/analysis/xisresp}}, which calls the tools {\hbox{\rm{\small XISRMFGEN}}} and {\hbox{\rm{\small XISSIMARFGEN}}}, was used with the ``medium'' input to obtain individual ancillary response files (arfs) and redistribution matrix files (rmfs). Finally, we combined the spectra and response files from the two front-illuminated instruments (XIS0 and XIS3) using the \ftool\ \addascaspec\ to increase signal-to-noise. The spectra, background and responses were then rebinned by a factor of eight to 512 channels using \rbnpha\ and \rbnrmf.  The \ftool\ \grppha\ was used to give at least 100 counts per spectral bin. Throughout this paper BI and FI spectra are fitted in the 0.5-8.0\kev\ and 0.7-10\kev\ energy range respectively. Unless stated otherwise, the 1.7--2.3\kev\ energy is ignored due to the  possible presence of uncalibrated instrumental features. All energy axes are in the observed frame.
 
For the hard X-ray detector (\hxd, \citealt{SUZ_HXD}) we again reprocessed the unfiltered event files following the data reduction guide (only the \pin\ data are used in this analyses). Since the \hxd\ is a collimating rather than an imaging instrument, estimating the background requires individual consideration of the non X-ray instrumental background (NXB) and cosmic X-ray background (CXB). The appropriate response file was downloaded\footnote{http://www.astro.isas.ac.jp/suzaku/analysis/hxd/} together with the tuned (Model D) background. A common good-time interval was obtained with \mgtime\ which combines the good time of the event and background file, and \xselect\ was used to extract spectral products. Dead time corrections were applied with \hxddtcor, and the exposures of the NXB spectra were increased by a factor of ten, as instructed by the data reduction guide. The contribution from the CXB was simulated using the form of \cite{Boldt87}, with the appropriate normalisation for the \hxd\ nominal pointing. The NXB and CXB spectra were then combined using \mathpha\ to give a total background spectrum, to which a 2~per~cent systematic uncertainty was added. The \pin\ spectrum was grouped to have a minimum of 500 counts per energy bin to improve statistics, and again allow the use of $\chisq$ minimization during spectral fitting. The \pin\ data reduction yielded a total source rate of $(3.27 \pm 0.02)\times10^{-1}$\ctps\ which is 38.5~per~cent of the total observed flux, with a good exposure time of $\approx234$\ks. We note here that since the main purpose of our analysis is to investigate the spectral variability of \n,  all \pin\ spectra presented in this paper are constrained to be strictly simultaneous with their \xis\ counterpart.  Furthermore, we restrict all our \pin\ analyses to the energy range 15.0--40.0\kev\ and fit it simultaneously with the \xis\ data by adding a normalization factor of 1.18{\footnote{See Paper~1 for a discussion of the effects in letting this cross normalization vary.}} with respect to that of the FI spectrum, as recommended in the \suzaku\ Data Reduction Guide.

\begin{figure}[!t]
\centering
{
 \rotatebox{270}{
\resizebox{!}{11.cm} 
{\includegraphics[clip=true]{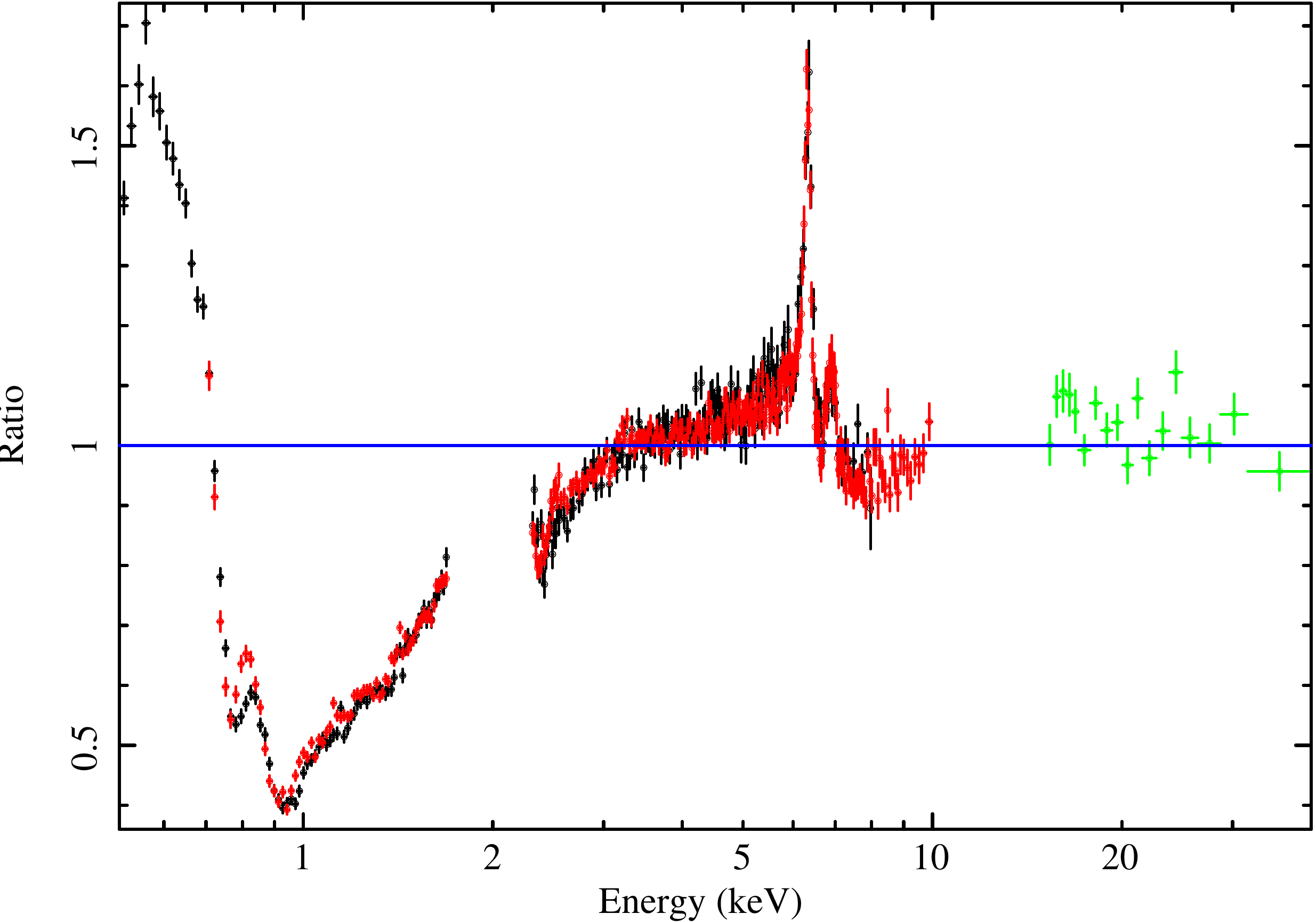}  
}}}
\caption{Time-averaged data/model ratio to an absorbed powerlaw. \xis\ back and front illuminated spectra are shown in black and red respectively. The \pin\ data are shown in green. A powerlaw was fitted in the 3.0--4.0 and 7.5--40.0\kev energy range, The residuals shows the various complexities including the presence of a strong warm absorber  seen at soft energies together with a superposition of emission lines at $\sim6.4$\kev.  }
\label{fig_ratio2pl}

\end{figure}

\section{Data Analyses and Results}
\label{analyses}

Fig.~\ref{fig_ratio2pl} shows the time-averaged \xis/\pin\ spectra for the entire data set compared to a simple absorbed powerlaw model 
(\nh\ frozen at the value of $9.91\times 10^{20}\pcmsq$ as found in Paper~1) fitted between 3.0--4.0\kev\ and 7.5--40.0\kev.  The presence of a strong warm absorber is clearly seen below 2\kev\ as is the presence of various emission features peaking at 6.4\kev. Detailed fits to the full \suzaku\ spectrum were presented in Paper~1. It was found that the time averaged spectrum was fitted well by a model consisting of a power-law continuum, blackbody-like soft excess, and X-ray reflection from both cold/neutral distant material and an ionized accretion disk; the observed emission from these components are also strongly affected by a three zone warm absorber.  In order to adequately describe the time-averaged spectrum, a fraction ($\sim$15\%) of the continuum emission scatters around or leaks through the warm absorbing gas.  See Paper~1 (Table~1) for a detailed discussion of the time-average spectral model.

In the following sections we will concentrate on the spectral variation of \n, initially using a number of model-independent techniques.

\begin{figure}[!t]
\centering
{
 \rotatebox{0}{ 
{\includegraphics[width=14cm]{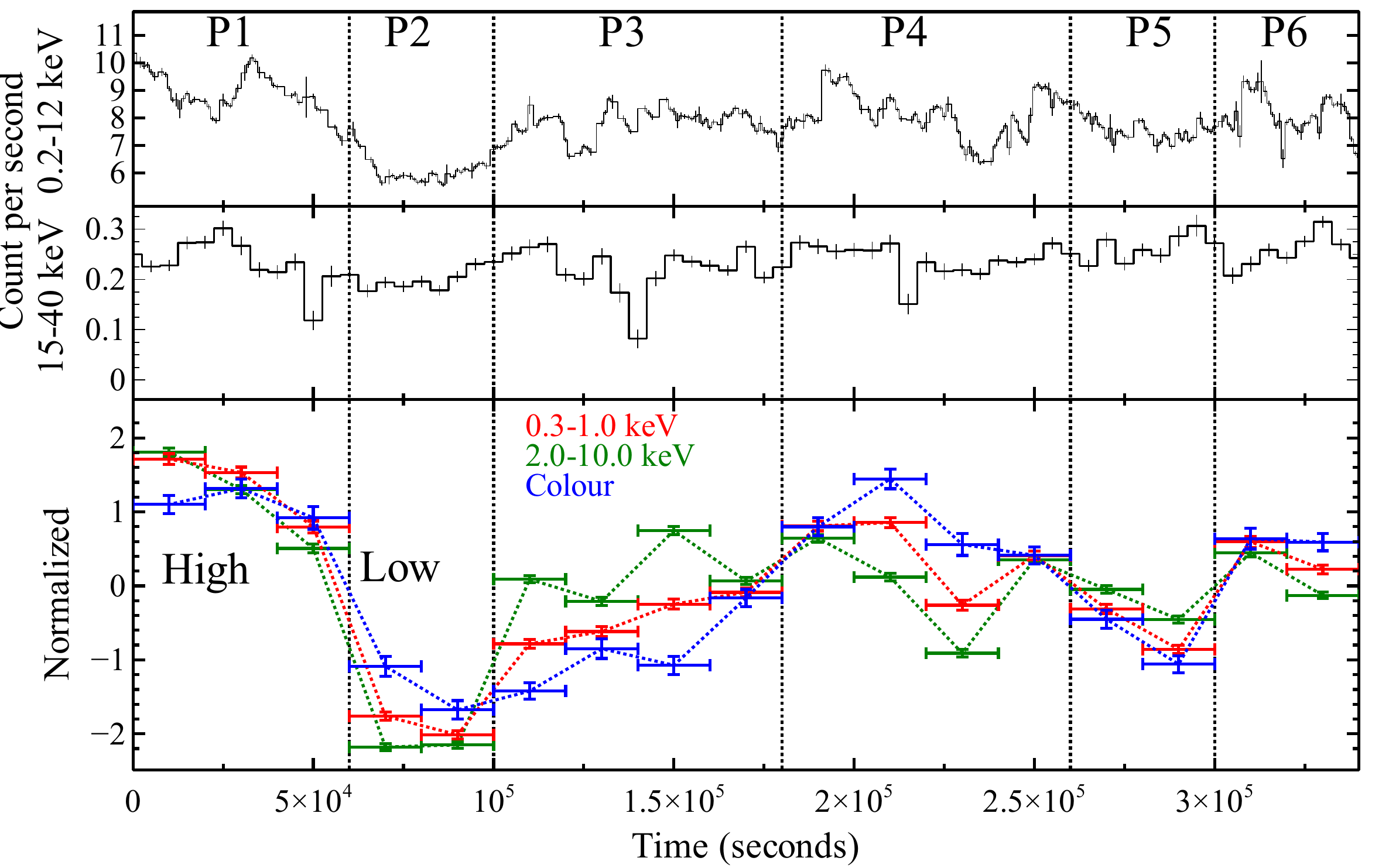}  
}}}
\vspace*{-0.2cm}
\caption{{\it Top:} 0.2--12.0\kev\ \xis\ light-curve in intervals of 1\ks. {\it Middle:} 15--40\kev\ \pin\ light-curve in intervals of 5\ks. {\it Bottom:} Normalised soft (0.3-1.0\kev, red), hard (2.0-10.0\kev, green) and soft/hard (colour, blue) count-rate as a function of time. In the lower panel these were normalised by dividing the difference between the count rate in each  20\ks\ time bin and mean count rate by the standard deviation.  For \S{\thinspace\ref{time_resolved}} the light-curve has been divided into six intervals (P1 to P6 indicated by the dotted vertical lines. We also indicate intervals where both the hard, soft, and softness ratio are at distinct high and low levels. }
\label{fig_lightcurve1}
\end{figure}

\subsection{Colour changes}
\label{variability}

It can be seen from the background-subtracted light curve in Fig.~\ref{fig_lightcurve1} (Top) that the X-ray continuum emission in \n\ shows clear flux variability on a variety of time scales. The light curve shown above for the \xis\ data has bin sizes of 1\ks\ and is the result of the combined data for the three \xis\  instruments. The \pin\ light curve in the 15--40\kev\ range is shown in the middle panel with bin sizes of 5\ks. In the bottom panel we show the normalised 0.3--1.0\kev\ (red; soft) and 2.0--10.0\kev\ (green; hard) count rate together with the colour (blue), defined here as the ratio between soft and hard count rate. In order to better show the evolution of the colour, soft and hard energy band, we have split the data into 20\ks\ time intervals. During the full observing campaign, \n\ initially dropped in flux by a factor of $\sim1.8$ in the first 100\ks\ and subsequently went back to an intermediate flux level during the latter $\sim200$\ks. When comparing the evolution of the soft, hard and softness ratio, we see that there are {\it approximately} six distinct periods (P1--P6) which have been highlighted in Fig.~\ref{fig_lightcurve1}. During the first and last two periods the softness ratio appears to follow both hard and soft normalised rate variation. However, this is not the case in periods 3 and 4 where the softness ratio is significantly below and above the two normalised flux bands, respectively. In order to characterise these spectral variations in a model-independent manner, we begin by applying the ``flux--flux''  analysis outlined in \cite{Taylor03}.

\subsection{Flux-flux analysis}
\label{flux_flux}

\cite{Taylor03} showed that for Seyfert galaxies, the flux measured in a specific energy band, $F(E)$, can be expressed as the sum of constant ($C_{\rm s}$ and $C_{\rm h}$) and variable $(F_{\rm s}$ and $ F_{\rm h}$) components, so that the flux in a hard energy band (subscript {\rm h}), can be related to that of a soft band (subscript {\rm s}),  by the expression: 
\begin{equation} \label{eq1}
F_{\rm h} = k(F_{\rm s} - C_{\rm s})^{\alpha} + C_{\rm h}, 
\end{equation}

If the varying component has a constant spectral shape, the flux-flux relation will be linear ($\alpha =1 $). If, however, the spectral variability is intrinsic to a single pivoting powerlaw with no contribution from any other component then plotting $F_{\rm h}$ versus $F_{\rm s}$ will result in a simple powerlaw of the form $F_{\rm h} = kF_{\rm s}^{\alpha}$ where $\alpha$ can be used to estimate the pivot energy (\citealt{zdziarskietal2003} and Fig.~1 of \citealt{Taylor03}).

\begin{figure}[!t]

\centering{
{\includegraphics[scale=0.42, clip=true]{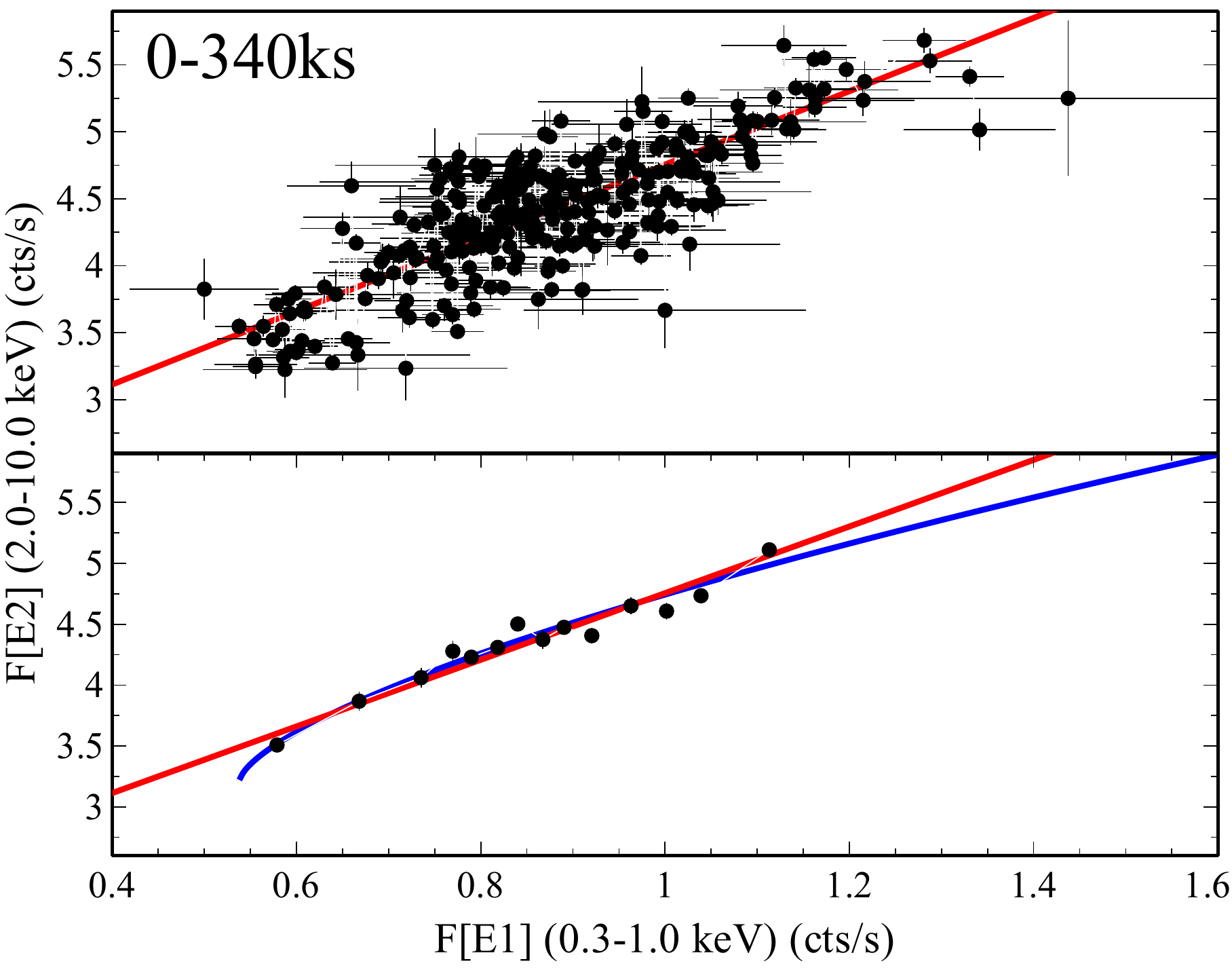}}\vspace*{0.15cm}
 
{\includegraphics[scale=0.42, clip=true]{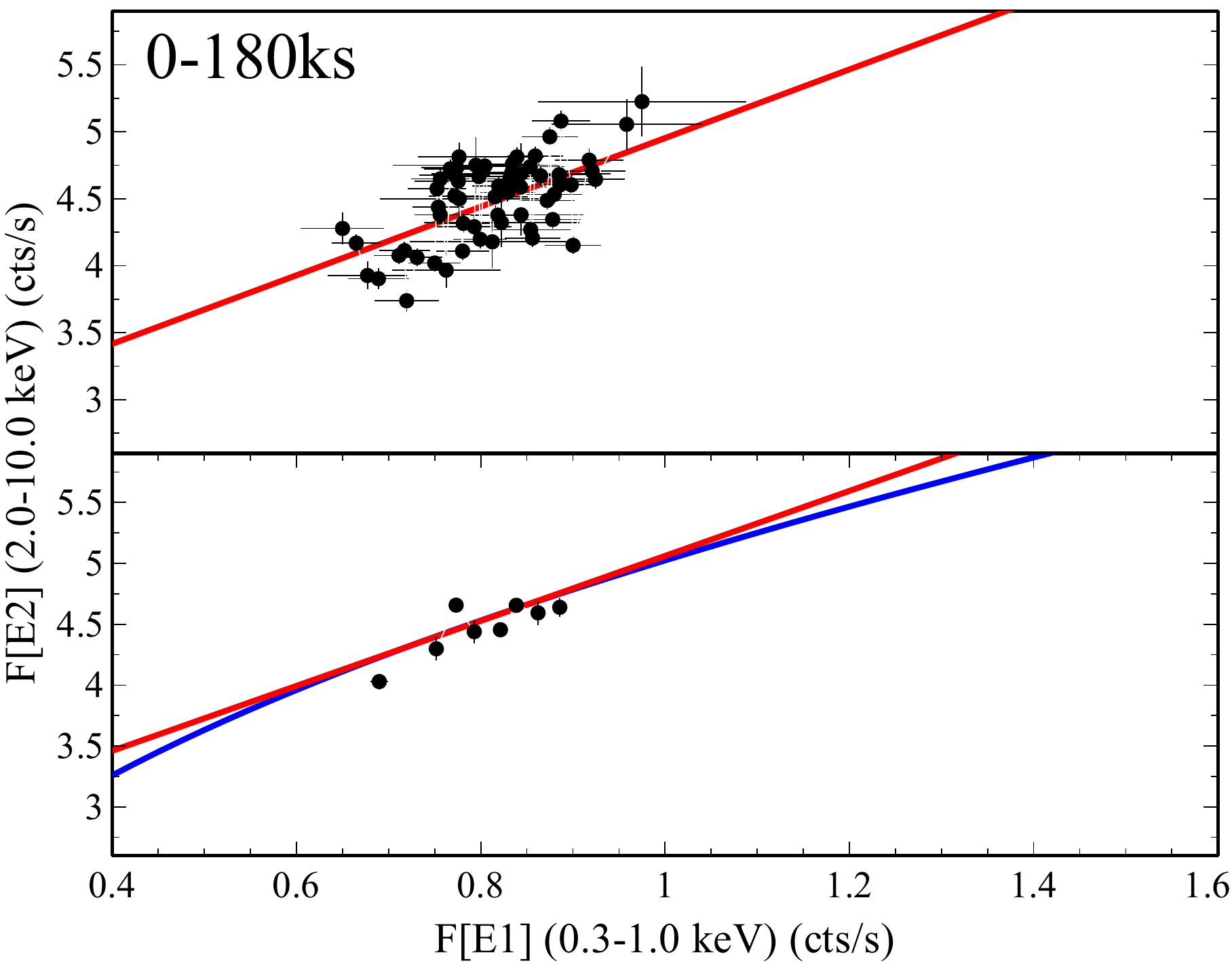}}
{\includegraphics[scale=0.42, clip=true]{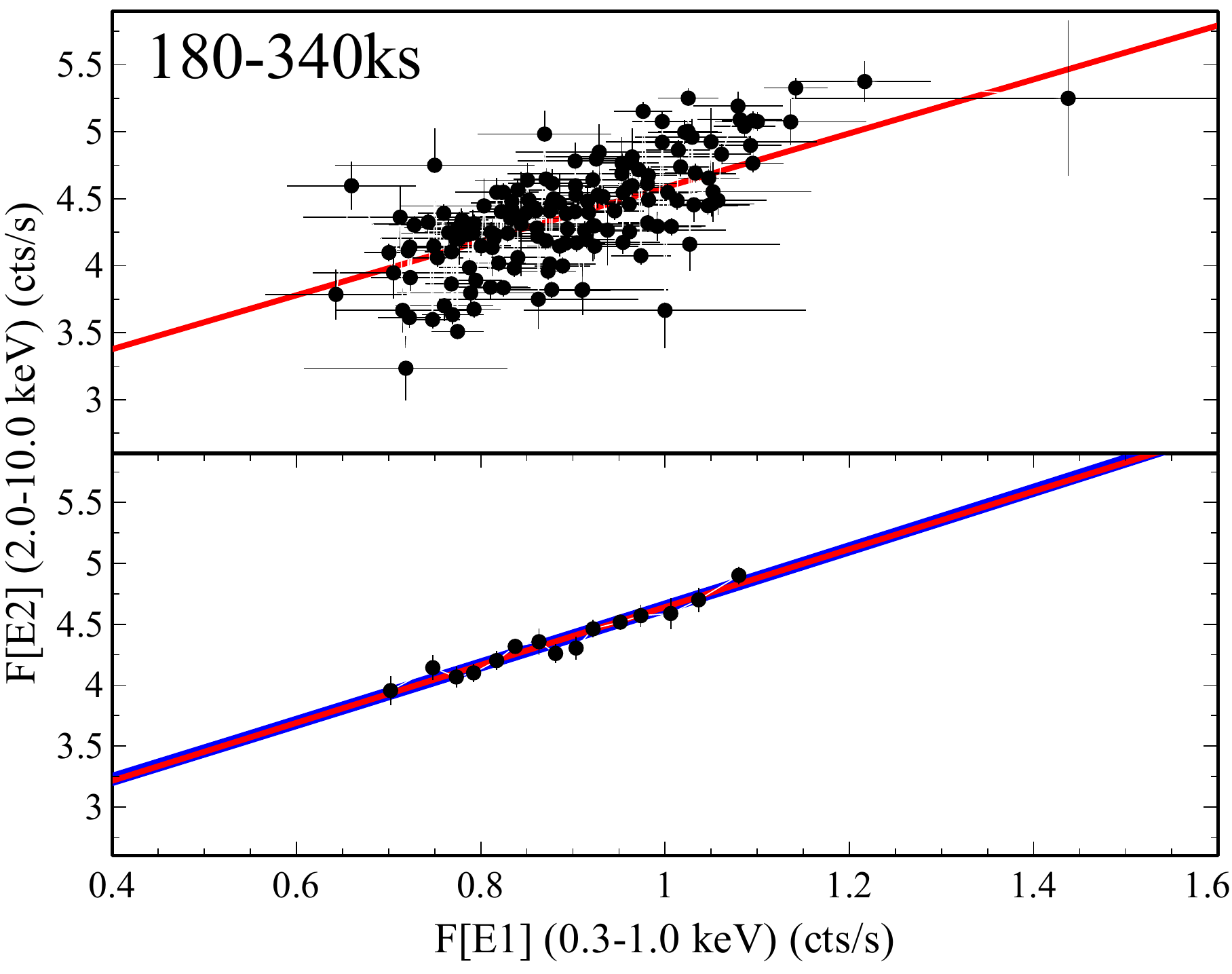}}
}
\vspace*{-0.35cm}
\caption{\label{fig_flux_flux1}Flux-flux plots for \n. Hard (2.0-10.0\kev) versus soft (0.3-1.0\kev) count rate.  {\it Top panels:} The points are from the light curve having 1\ks\ time bins. {\it Bottom panels}: Same as above but after averaging over the points and adding 2~per~cent systematic to each XIS count. In all cases, the red line shows the best linear fit to the data and the blue curve represents the best fit described by Equation~1. }
\end{figure}

Figure~\ref{fig_flux_flux1} (Top) shows the hard (2.0--10.0\kev) versus soft (0.3--1.0\kev) count rates (flux--flux plot) for \n\ from the 1\ks\ binned \xis\ light-curves. The scatter is partly due to poor statistics and is mostly removed when  we bin the data. The red line is the best linear model which is summarised in Table~\ref{table1}. Although visually the linear relationship appears correct, the fit is not statistically acceptable, with $\chisq/\nu =40.4/12$. When we add a systematic error of 2~per~cent to each XIS count (as per \citealt{nado2011}), the resulting fit gives $\chisq/\nu =11.5/12$. This extra systematic is likely due to uncertainties in CCD contamination (\citealt{nado2011}). Replacing the linear relation with equation~\ref{eq1} results in only a mild improvement ($\chisq/\nu =8.2/10$; blue curve). This best fit cannot rule out weak pivoting nor does it statistically require it ($\alpha = 0.63\pm 0.16$ with constant offsets on the hard and soft axes of $C_{\rm h} = 3.2\pm0.2$ and $C_{\rm s} = 0.54\pm0.02$ respectively; 1$\sigma$). According to Figure~1 of 
\citet{Taylor03}, $\alpha$ ranging from approximately 0.55 to 0.8 implies a powerlaw with a pivot energy ranging from 30 to 300\kev. In Section~\ref{time_resolved} we show using time-resolved spectra, evidence for a distinct change in the spectrum between the first and second half of the observation. In order to investigate whether any change is apparent in the flux-flux plot between these two time intervals, we created flux-flux plots for the first 180\ks\ (P1+P2+P3 in Fig.~\ref{fig_lightcurve1}) and the latter 160\ks\ (P4+P5+P6). These are shown Figure~\ref{fig_flux_flux1} (Bottom left and right respectively) and the fits are summarised  in Table~\ref{table1}.

\begin{table}[!bt]
\caption[]{\label{table1}Summary of linear and powerlaw fits to the flux-flux plots shown in Fig~\ref{fig_flux_flux1}. All errors are $1\sigma$. $^a$ The powerlaw is of the type shown in Equation~\ref{eq1}.}
\vspace*{-0.3cm}
{%
\newcommand{\mc}[3]{\multicolumn{#1}{#2}{#3}}
\begin{center}
 \begin{tabular}{lcllcll}
\hline
Period & \mc{3}{c}{Linear model} & \mc{3}{c}{Powerlaw model$^a$}\\
× & $k_{\rm lin}$ & \mc{1}{c}{C} & \mc{1}{c}{$\chisq$/d.o.f} & $k_{\rm pow}$ & \mc{1}{c}{$\alpha$} & \mc{1}{c}{$\chisq$/d.o.f.}\\\hline
Full range & $2.7\pm0.2$ & \mc{1}{c}{$2.1\pm0.2$} & \mc{1}{c}{11.5/12} & $2.5\pm0.2$ & \mc{1}{c}{$0.63\pm0.16$} & \mc{1}{c}{8.2/10}\\
0-180ks & $2.96\pm0.86$ & \mc{1}{c}{$2.1\pm0.7$} & \mc{1}{c}{9.9/6} & $4.3\pm0.3$ & \mc{1}{c}{$0.55\pm0.22$}  & \mc{1}{c}{9.5/4}\\
180-340ks & $2.4\pm0.2$ & \mc{1}{c}{$2.3\pm0.1$} & \mc{1}{c}{7.3/13} & $2.2\pm0.2$ & \mc{1}{c}{$1.05\pm0.07$} & \mc{1}{c}{7.2/11}\\\hline
\end{tabular}
\end{center}

}%

\end{table}

The total flux range probed in this observation is quite modest ($\approx 1.8$ between highest and lowest flux bins) compared with other Seyfert~1s such as MCG-6-30-15 (factor of 3; \citealt{Vaug04}) or NGC~4051 (factor of 8 over longer time scales; \citealt{Uttley04}). This limited flux range prevents strong constraints on the derived value of $\alpha$, as is clear from the results shown in Table~\ref{table1}.

\begin{figure}[!t]
\centering
{
 \rotatebox{0}{
{\includegraphics[scale=0.52, clip=true]{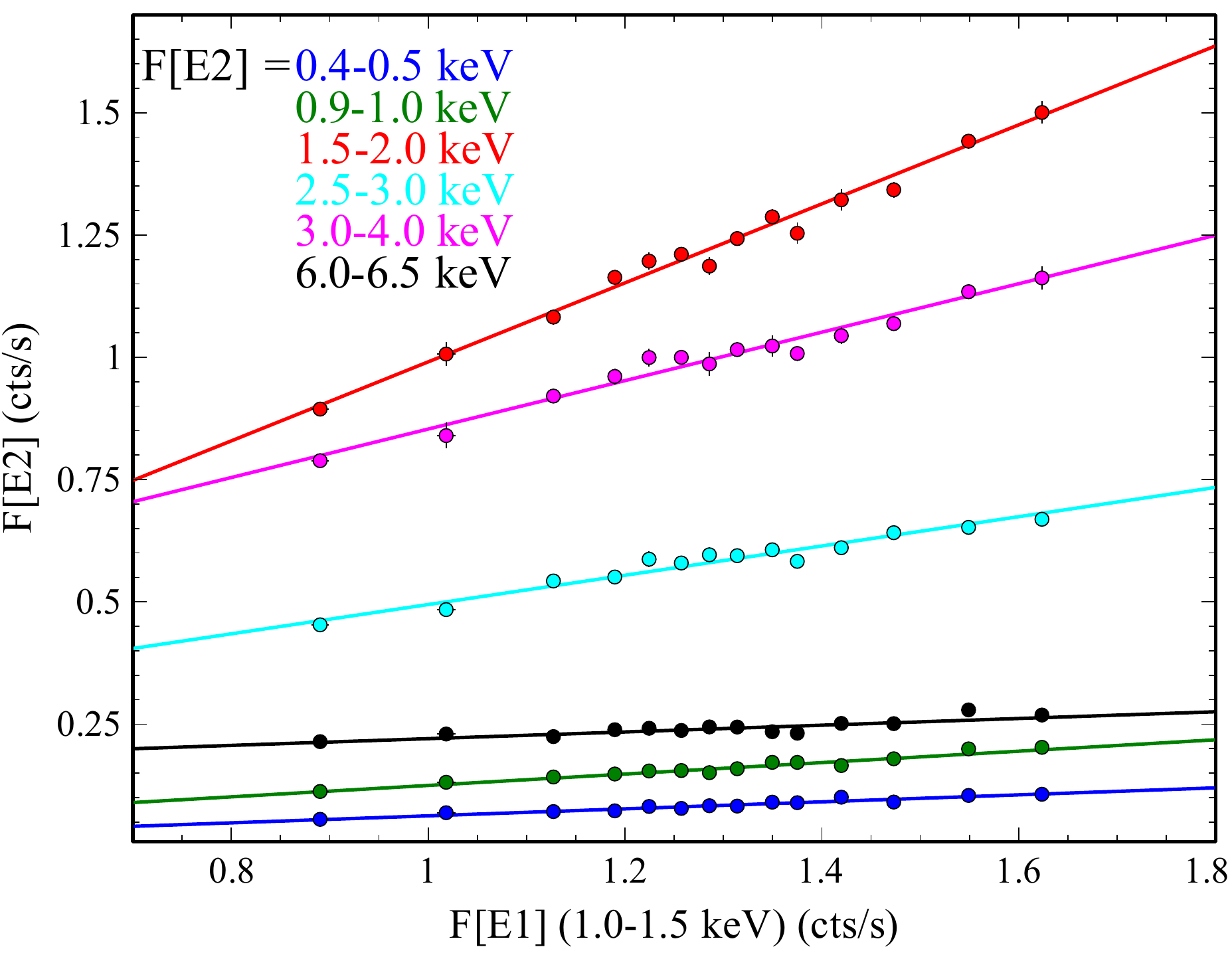}}
}}
\vspace*{-0.2cm}
\caption{\label{fig_flux_flux2} Binned band flux F({\rm E2}) versus 1.0--1.5\kev\ flux. The various lines shows the best linear-fit to a sample of energy range. By obtaining the constant offset for the various energies with respect to a standard range we can deduce the shape of the constant component (\S{\thinspace\ref{flux_flux}}; see Fig.~\ref{fig_flux_flux3}).}
\end{figure}

Under the assumption that the spectral variability in \n\ can be described by the sum of a constant and variable component where the variable component has a constant shape, i.e. $\alpha = 1$ in equation~1, we can use flux-flux plots for a variety of energy bands with respect to a single reference band (Fig.~\ref{fig_flux_flux2}) to obtain the spectral shape of the constant component. Since the best fits are indeed roughly consistent with $\alpha=1$, this assumption is probably valid.  \citet{Taylor03} and \citet{Vaug04} describes the methodology used here in detail.

\begin{figure}[!t]
\centering
{
 \rotatebox{0}{
{\includegraphics[scale=0.5, clip=true]{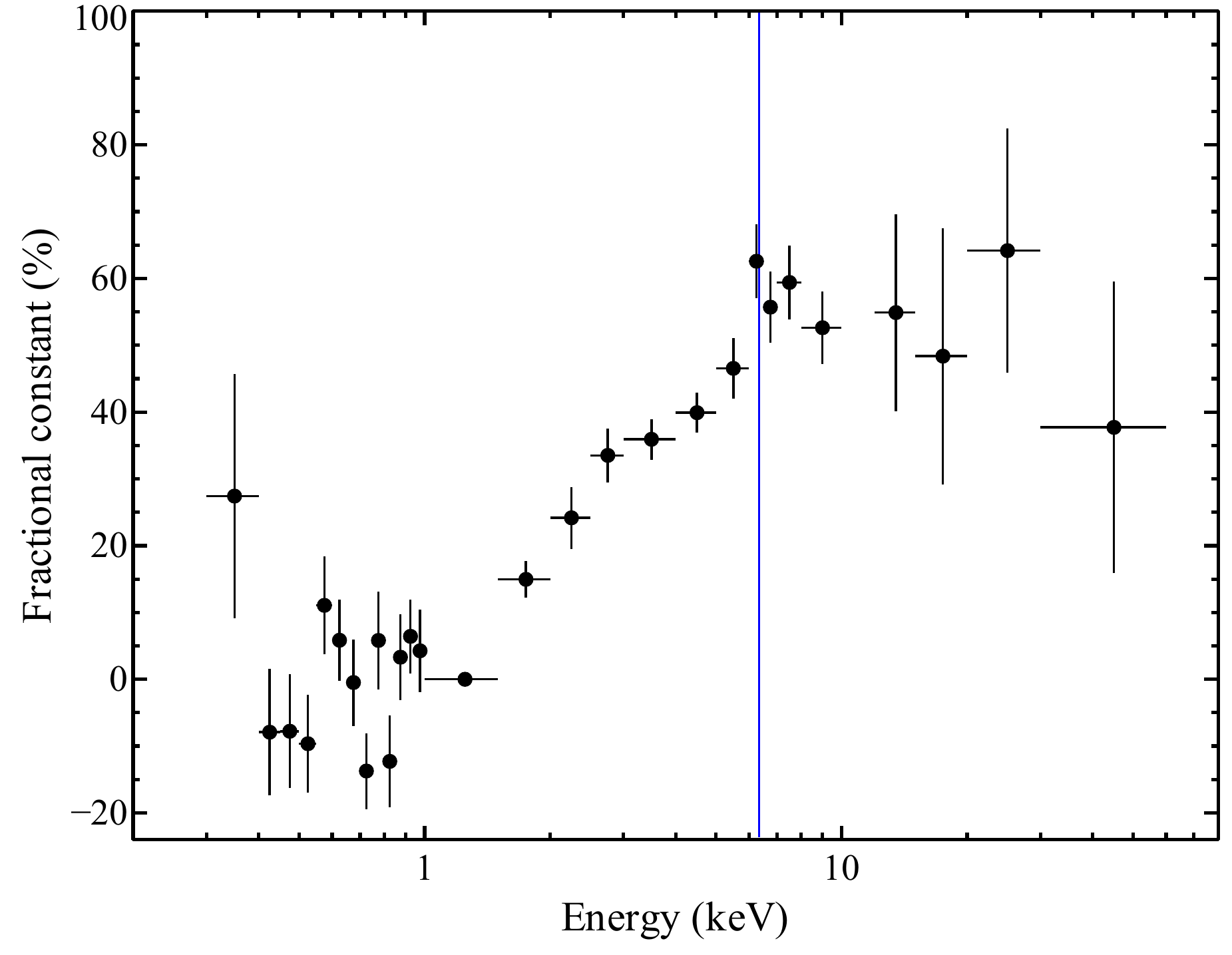}  
}}}
\vspace*{-0.3cm}
\caption{ Shape of the constant component as obtained from the linear flux--flux relations shown in Fig.~\ref{fig_flux_flux2}. The constant vertical offsets were divided by the flux in each energy bin so as to show the shape of the constant component independent of the instrumental response. The vertical line shows the expected position of the iron~\ka\ emission line at 6.4\kev\ (rest frame).  }
\label{fig_flux_flux3}
\end{figure}

On inspection, the reference band, F(E1), is chosen to be the 1.0--1.5\kev\ energy range as this shows the least contribution from the constant component (see Fig.~\ref{fig_flux_flux3} and Fig.~14).  By measuring the offset of the flux--flux plot as a function of the discrete energy band $c(E)$, we can estimate the fractional contribution of the constant component to the spectrum as a function of energy.  Fig.~\ref{fig_flux_flux3} shows the offset $c$ as a function of energy. The offset was normalised relative to the average flux in each specific energy band. This normalisation is applied so as to show the shape of the constant component independent of the instrument response.  Note that by virtue of choosing the 1.0--1.5\kev\ as the reference band, we are assuming that this band contributes little to the constant component, i.e $c(1.0-1.5$\kev)$~\approx0$. If, on the other hand, there is indeed a non-negligible contribution from the constant component to the overall spectrum at this energy range, then we should expect the scale to increase. However, the overall shape of the constant component will remain the same.

From Fig.~\ref{fig_flux_flux3} it can be seen that the constant component is significantly harder than the total spectrum in the 2.0-10.0\kev\ band. A similar analysis for \m\ (\citealt{Vaug04}; see their Fig.~11) shows that the constant component in \m\ has an increasingly larger amplitude as the energy softens below 1\kev. This is not the case for \n\ where the constant component seems to flatten in this energy range. Furthermore we note the presence of a feature at around 6\kev\ (indicated by the solid vertical line in Fig.~\ref{fig_flux_flux3}) in the spectrum of the constant component as well as a further flattening of the spectrum above 10\kev. These features are qualitatively very similar to that found for \m\ by Taylor et al. (\citeyear{Taylor03}; see their Fig.~5) where the authors identified the constant component as reflection from the inner accretion disk, similar to what was  proposed by \cite{Fabianvaughan03}.

It should be stressed that the ``constant component'' identified via this model independent method and shown in Fig.~\ref{fig_flux_flux3}, needs not be a single component nor does it need to be truly constant. It is plausible that this component be in fact the  superposition of two distinct components whose variability are significantly weaker than that of the ``variable''  component.

% CSR : new paragraph
As shown in Paper~1, the time-averaged spectrum of \n\ displays a fairly strong narrow 6.4\kev\ emission line; this is a strong indication for X-ray reflection from cold and distant material. While the exact location of this material is unclear, the limits on the width of the line ($FWHM<5500\kmps$; \citealt{Shu2010}) under the assumption of dynamical equilibrium,  place this cold material at a distance of $r>10^4r_g$ from the black hole.    For a mass of $M=3\times 10^7\msun$ (\citet{VestergaardPeterson2006} quotes values ranging from  $M=(2.8-3.0) \times 10^7\msun$ based on optical single-epoch spectroscopy), this places the cold reflector at $r>4.4\times 10^{16}\cm$ which corresponds to 17 light days or more.   Thus, during the span of this observation, it is expected that the distant/cold reflection component will be approximately constant. Some, but not all, of the constant component revealed in Fig.~\ref{fig_flux_flux3} can be identified with this distant reflection.

\begin{figure}[!t]
\centering
{
 \rotatebox{0}{
{\includegraphics[width=10.0cm]{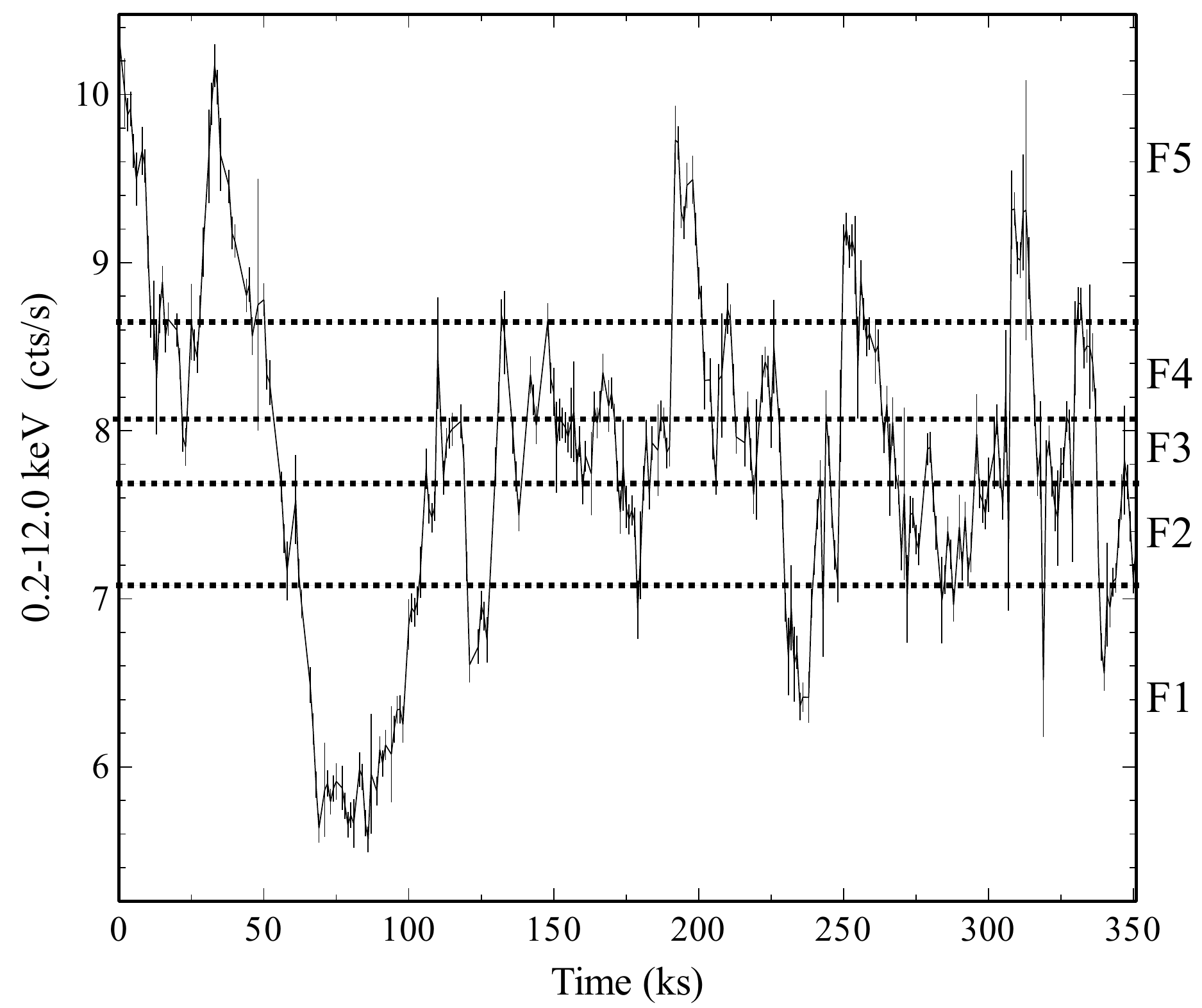}  
}}}
\vspace*{-0.2cm}
\caption{ 0.3-12\kev\ light curve (1\ks\ bin) showing the various flux bins used in the following section.   }
\label{fig_light_flux}
\end{figure}

\subsection{Flux resolved spectroscopy}
\label{flux_resolved}

\begin{figure}[!t]
\centering
{
 \rotatebox{270}{
\resizebox{!}{12cm} 
{\includegraphics{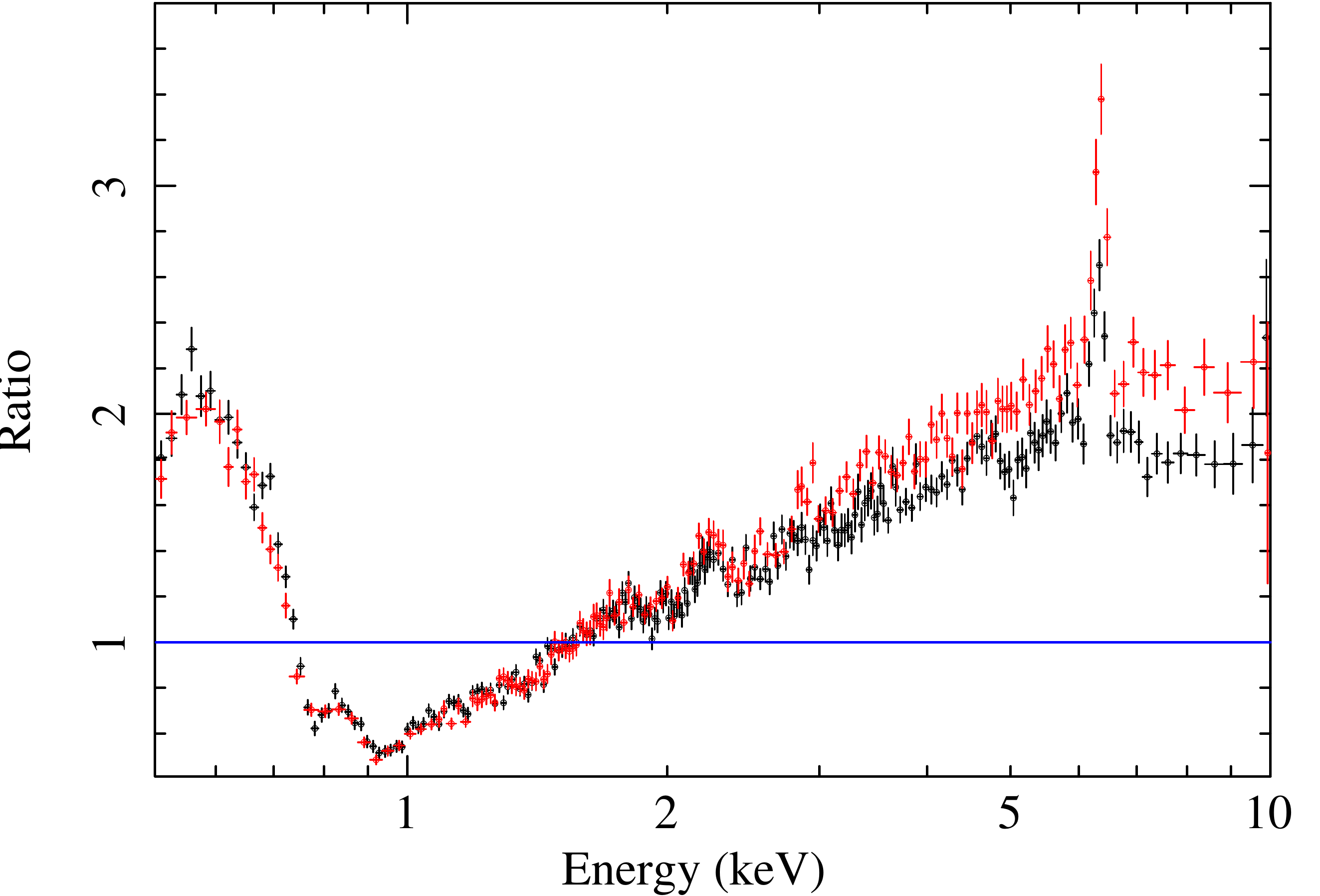}  
}}}
\caption{ Ratio of high (F5; black) and low (F1; red) flux BI spectrum to an absorbed ($\Gamma=1.8$) powerlaw. The shape of the soft energy range is remarkably similar in both cases suggesting that the warm absorber is not changing. As is usually seen in other sources (e.g \m) the overall spectrum becomes harder with a more prominent iron feature at low fluxes. This is as expected from the two-component model described in \S~\ref{flux_flux}.  }
\label{fig_high_low_flux_ratio2pl}
\end{figure}

A further way to explore the X-ray variability is by examining the evolution of the energy spectra with flux. To do this we divided the data into five flux intervals (F1--F5 shown in Fig.~\ref{fig_light_flux}) each having approximately the same number of photons ($\sim 2.2\times10^5$ and $1.4\times10^5$ for the BI and FI respectively). Spectra and responses were produced for each flux interval. To illustrate the spectral change as a function of flux we show in Fig.~\ref{fig_high_low_flux_ratio2pl} both the high (F5; black) and low (F1; red) spectra as ratio to an absorbed (\nh$ = 9.91\times 10^{20}\pcmsq $) powerlaw  having $\Gamma = 1.8$. It is clear that at low flux the spectrum is harder above 2\kev\ which is as expected from the two-component model described in the previous section. Assuming the spectrum is composed of a soft-variable component together with a hard-constant component (similar to that shown in Fig.~\ref{fig_flux_flux3}) then as the total flux decreases, the relative contribution of the harder component increases. This behaviour is also seen in other Seyfert galaxies such as \m\ (\citealt{Vaug04}, and references therein).
 
 \vspace*{-0.5cm}
\subsubsection{Difference spectra}

In this section we use the spectra obtained from the five flux levels described above to produce flux-resolved ``difference-spectra''. By subtracting the lowest flux spectrum (F1) from the remaining spectra we effectively remove the contribution of any constant (emission) component. This technique was described in detail in \cite{Fabianvaughan03} and results in a spectrum consisting of the variable component modified by any intervening (Galactic and intrinsic) absorption.

\begin{figure}[!t]
\centering{
\mbox{\hspace{-1.cm}\subfigure{\includegraphics[width=3.1in]{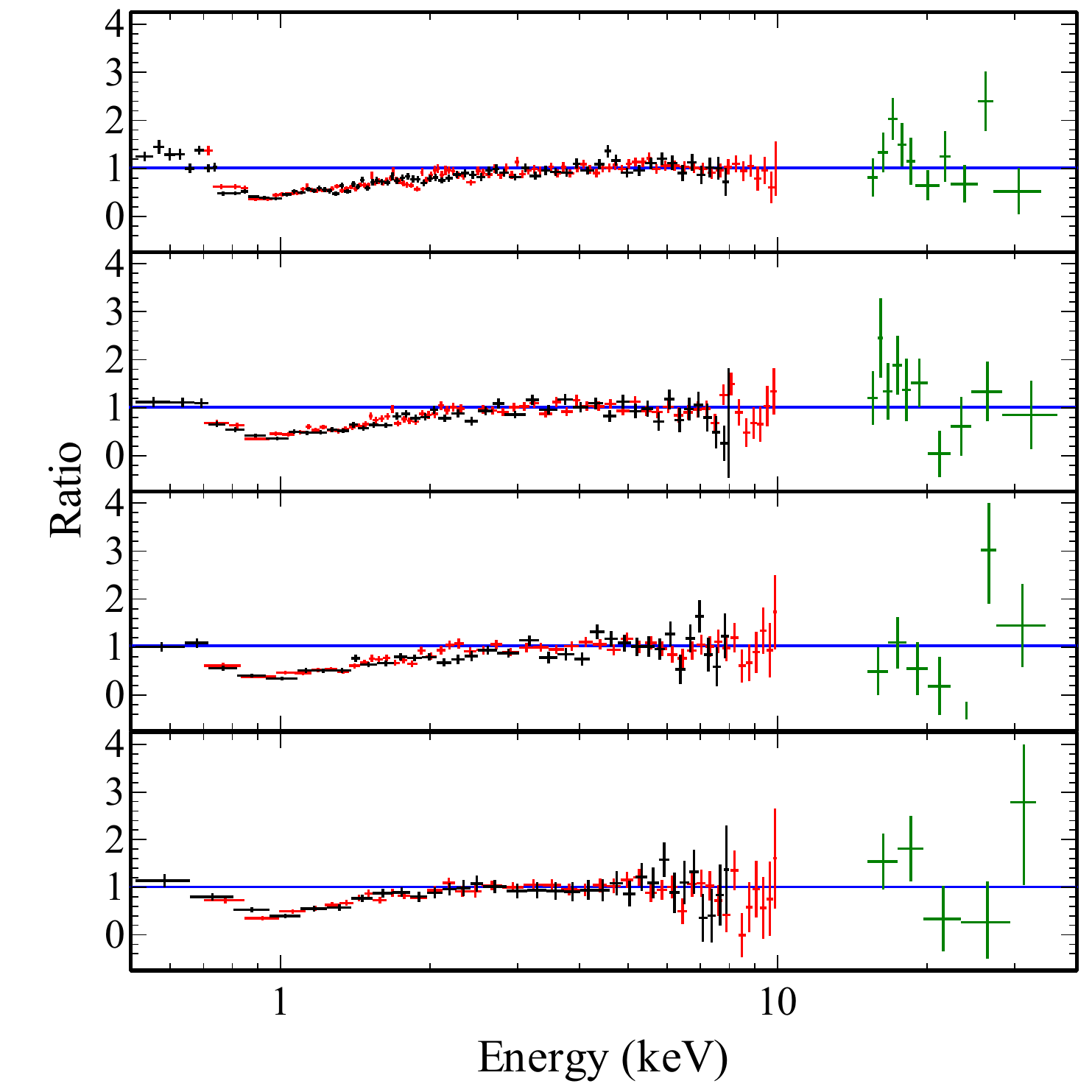}
\quad
\subfigure{\includegraphics[width=3.1in]{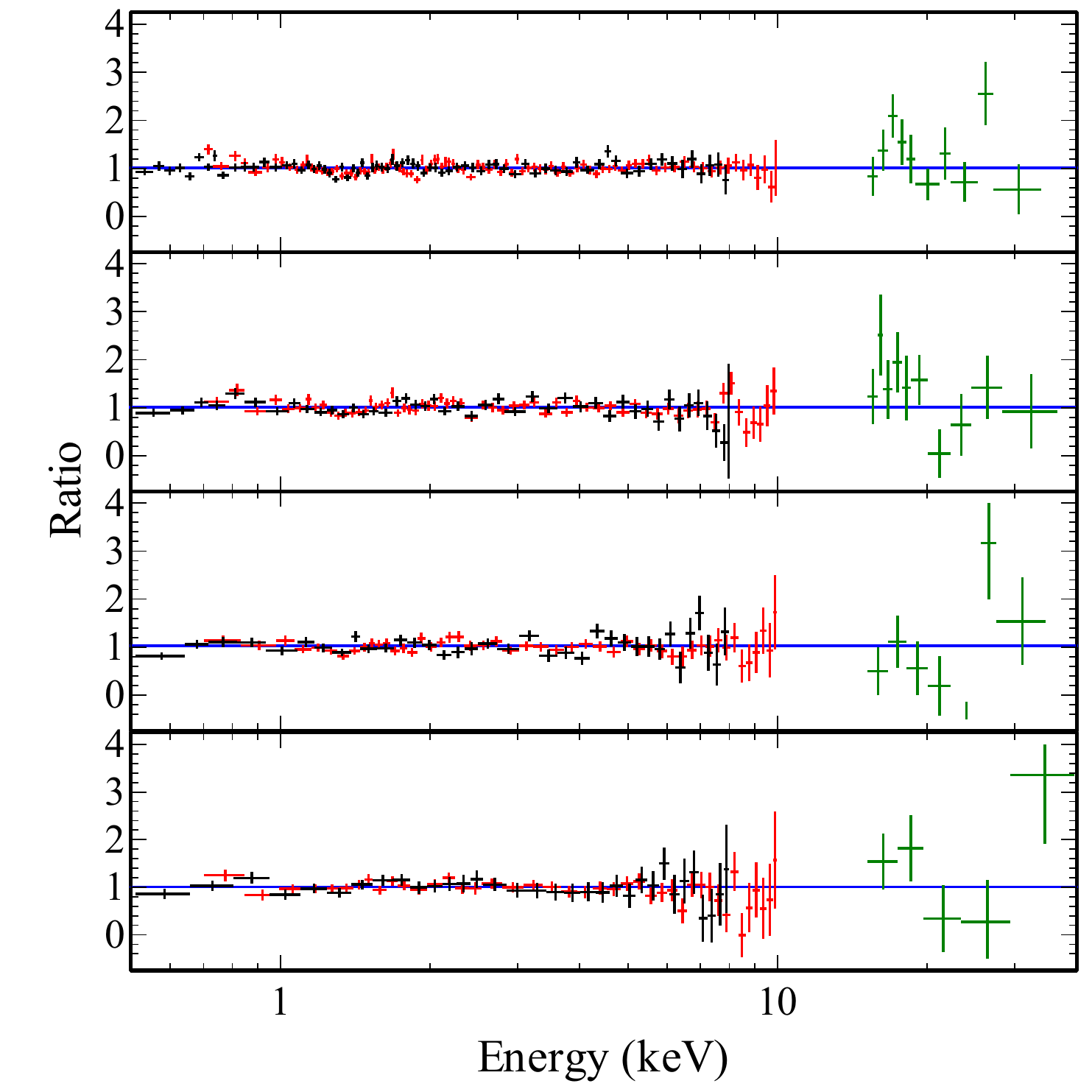} }}}}
\vspace*{-0.6cm}
\caption{ {\it Left:} Data/model ratio of the flux-resolved difference-spectra to an absorbed powerlaw. The difference-spectra shown are (from top to bottom) F5, F4, F3, and F2 minus F1. The spectra were fit above 3\kev\ with a powerlaw and extended to the full energy range after the inclusion of neutral Galactic absorption. BI, FI and PIN data are shown in black, red and green respectively. {\it Right:} Now with the inclusion of a three-zone warm absorber. }
 \label{fig_dif_flux_pl}
 \end{figure}

In subtracting the \xis\ spectra we assume that the background for each period is constant and therefore use the difference-spectrum without background subtraction. To test this assumption we investigated the background in the \xis-FI data  over the five periods described in \S~\ref{time_resolved}. In each period, the 0.5--10 background count rate are found to be approximately $3\times 10^{-2} \ctsps$, which, in all cases, corresponds to less that 1.5~per~cent of the total source rate. Similarly, the 8--10\kev\ background count rate, although slightly more variable (ranging from $2.3\times 10^{-3}\ctsps$ to $2.7\times 10^{-3}\ctsps$), always remains at less than 5~per~cent of the total source counts. However, due to the strong time dependency of the \pin\ background, we produced background-subtracted spectra for each period and then created the difference-spectra using \mathpha, propagating the errors in each arithmetic operation. Figure~\ref{fig_dif_flux_pl} (Left) shows the data/model ratio for the four independent flux-resolved difference-spectra fitted with an absorbed powerlaw above 3\kev. Such an absorbed powerlaw results in a satisfactory fit with $\chisq/\nu =  1716.5/1669~(1.03)$ above 3\kev, with $\Gamma=1.86\pm0.05$ at the 90 per cent level of confidence. Allowing $\Gamma$ to vary between the various difference-spectra did not result in a statistically significant improvement to the overall fit, with a decrease in $\chisq$ of 7.8 for three degrees of freedom (F-test value of 2.5). Table~\ref{table2} details the values found for $\Gamma$ for the four difference-spectra as well as the global value shown above.
When we extend the data to cover the full range, we see clear flux deficit below 2\kev\ due to the warm absorber, which is not removed  in the difference-spectra.

\begin{table}[!ht]
\begin{center}
\caption[]{\label{table2}Summary of fits to the various flux-resolved difference-spectra. }
\vspace*{0.1cm}
{%
\newcommand{\mc}[3]{\multicolumn{#1}{#2}{#3}}
\begin{center}
\begin{tabular}{l|llll}
\hline
\hline
× & \mc{4}{c}{Flux-Resolved Difference Spectra}\\
\hline
Flux Bin$^a$ & \mc{2}{c}{$>3$\kev$^b$} & \mc{2}{c}{Full range$^c$}\\
\hline
× & \mc{1}{c}{Gamma Untied} & \mc{1}{c}{Gamma Tied} & \mc{1}{c}{Gamma Tied}& \mc{1}{c}{Gamma Tied }\\
F5 - F1 & \mc{1}{c}{$1.8\pm0.1$} & \mc{1}{c}{$1.86\pm0.05$} & \mc{1}{c}{$2.00\pm0.06$}&\mc{1}{c}{---}\\
F4 - F1 & \mc{1}{c}{$1.95\pm0.1$} & \mc{1}{c}{---} &\mc{1}{c}{---} &\mc{1}{c}{---}\\
F3 - F1 & \mc{1}{c}{$1.92\pm0.13$} & \mc{1}{c}{---} &\mc{1}{c}{---}& \mc{1}{c}{---}\\
F2 - F1 & \mc{1}{c}{$2.0\pm0.2$} & \mc{1}{c}{---} &\mc{1}{c}{---}& \mc{1}{c}{---}\\
T - B$^d$ & \mc{1}{c}{---} & \mc{1}{c}{---} &\mc{1}{c}{---} &\mc{1}{c}{$1.94\pm0.05$}\\
$\chisq/\nu$ & 1708.7/1666 (1.03) & 1716.5/1669 (1.03) &2995.3/2853 (1.05) & 753.6/709 (1.06)\\
\hline
\hline
\end{tabular}
\end{center}
}%
\end{center}

{\small Notes.- $^a$ The flux bins refers to the various flux slices shown in Fig.~\ref{fig_light_flux}.$^b$ The data were fit with a simple absorbed powerlaw in the 3--40\kev\ energy range. $^c$The full 0.5--40\kev\ energy range was fit after the inclusion of a three-zone warm absorber. $^d$ T and B refers to flux bins F5 + F4 and F1 + F2 respectively. All errors refers to the 90~per~cent confidence range. }

\end{table}

\begin{figure}[!t]
\centering
{
 \rotatebox{0}{
\resizebox{!}{8.cm} 
{\includegraphics{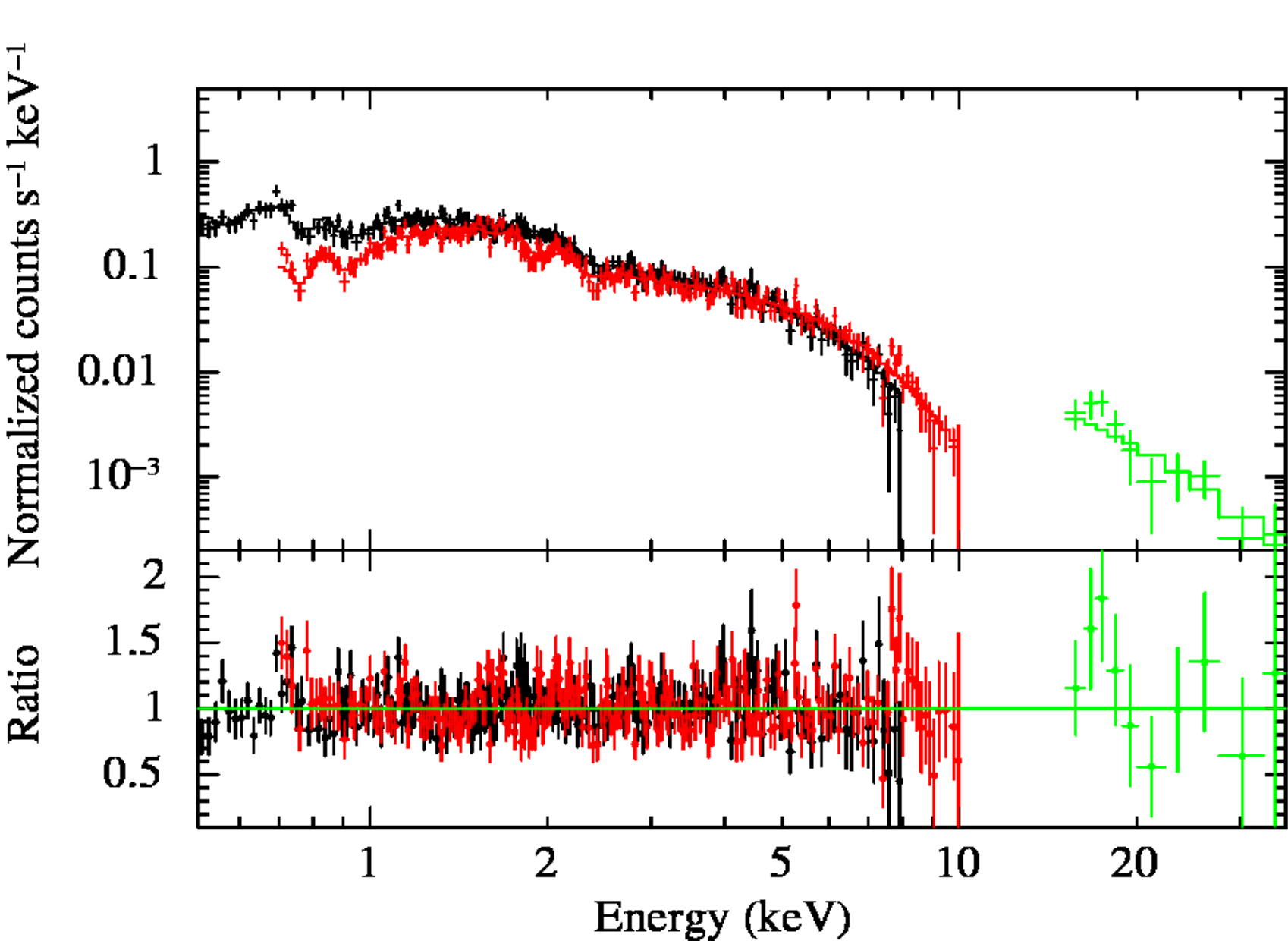}  
}}}
\vspace*{-0.3cm}
\caption{ Similar model to Fig.~\ref{fig_dif_flux_pl} (right) fitted to the difference-spectra resulting from the subtraction of the bottom two flux bins from the top two flux bins (see Fig.~\ref{fig_light_flux}). }
\label{fig_dif_flux_bottom2}
\end{figure}

In order to model the warm absorber we use the same photoionisation model as in Paper~1 (generated using \xstar v2.2.0). The free parameters of this absorption model are the column density and ionisation parameter of the absorbing medium. Following from the results of the time-averaged spectrum presented in Paper~1 we have modelled the warm absorber assuming three distinct zones. To reduce computational time we have frozen the ionization parameter of the various zones to the values found in Paper~1. We have also investigated varying the ionisation parameters whilst fixing the column densities to the values reported in Paper~1, however it was found to give similar results and so we chose to carry on with the former approach.  Because the \suzaku/\xis\ detectors do not have the spectral resolution capable of constraining the outflow velocities of the various warm absorbers zones, we follow the practice employed in Paper~1 and hold the redshifts of these components fixed at the cosmological value for \n. As such, in each period the free parameters of the model were the normalization of the powerlaw together with the column densities for the three absorbing zones. Figure~\ref{fig_dif_flux_pl} (Right) shows the data/model ratio for the various difference-spectra after the inclusion of the three-zone warm absorber.  The best fit resulted in $\chisq/\nu =  2995.3/2853~(1.05)$. Figure~\ref{fig_dif_flux_bottom2}  shows the fit to the difference-spectra resulting from the subtraction of the bottom two flux bins from the top two flux bins (see Fig.~\ref{fig_light_flux}). A simple powerlaw ($\Gamma=1.94\pm0.05$) modified by a three-zone warm absorber again results in an excellent fit to the difference-spectra ($\chisq/\nu =  753.6/709~(1.06)$). This fit to the full energy spectrum is also summarised in Table~\ref{table2}. The analysis presented in this section  re-enforces the result of the flux--flux analysis (\S~\ref{flux_flux})  that the shape of the variable component (the powerlaw in this case) does not vary significantly with flux.

\begin{figure}[!h]
\centering
{
 \rotatebox{0}{\hspace*{-0.4cm}
{\includegraphics[scale=0.53]{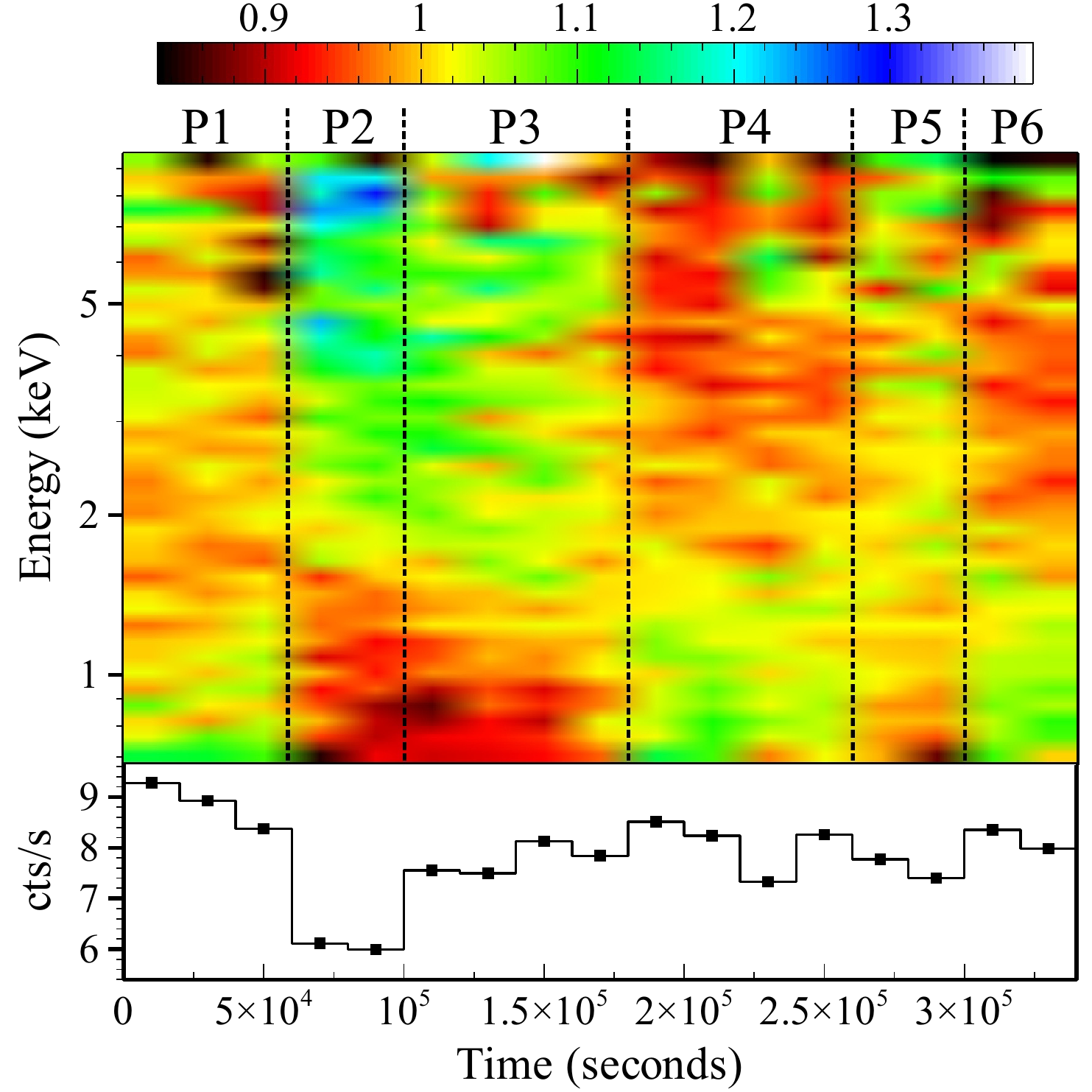}}
{\includegraphics[scale=0.53]{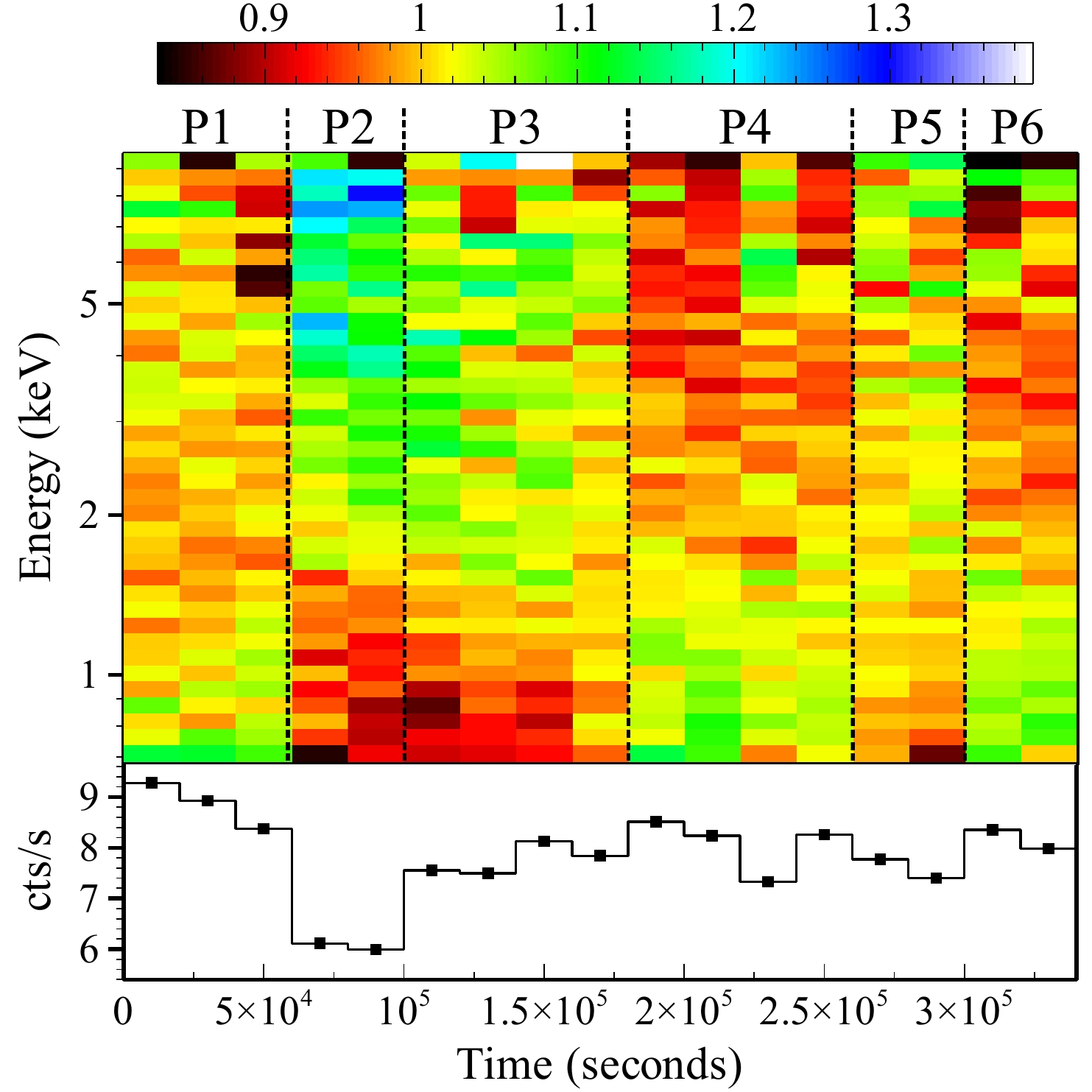}}
}}
\vspace*{-0.5cm}
\caption{Smoothed (left) and un-smoothed (right) time-resolved spectra for 20\ks\ segments. {\it Top:} 2-dimension image showing the ratio of the various spectra to the time-averaged spectrum as a function of energy and time. The typical standard deviation is approximately 0.03.  {\it Bottom:} Corresponding \xis\ count-rate. The vertical dotted lines show the position of possible changes in the spectral shape as compared to the average. Based on this image and Fig.~\ref{fig_lightcurve1} we divide the total observation into six periods.  }
\label{fig_image}
\end{figure}

\subsection{Time resolved spectroscopy}
\label{time_resolved}

In order to further visualise the variability we divided the full observation into seventeen 20\ks\ segments (real time) and extracted spectra and responses for each time segment. Each spectrum was then compared to the time-averaged spectrum by taking the ratio between them. The result was the ratio -- at all energies -- between the time-averaged spectrum and each 20\ks\ segment. Figure~\ref{fig_image} shows this ratio as a function of time and energy (Top), together with the total count rate for each segment (Bottom). Apparent changes to the overall spectrum are seen at approximately 60, 180 and 260\ks, with some indications of a further discontinuity at 300\ks. We note here that from Fig.~\ref{fig_image} it appears possible that there are narrow regions showing spectral variability at timescales shorter than 20\ks\ (e.g. halfway through the fourth period the colour code  between 5--7\kev\ changed  from orange to green implying a significance greater than $3\sigma$), however, spectra extracted at these timescales have very low signal-to-noise which prevents us from performing meaningful spectral decomposition. We stress that  this image is employed only to show the general trend in the energy-time parameter space. Based on the broad trends apparent in figure~\ref{fig_image} and the light curve shown in Fig.~\ref{fig_lightcurve1} we break the full observation into six distinct spectral periods (P1 to P6) in order to carry out a similar analysis to that of \S~\ref{flux_resolved} and produce difference-spectra for the five time periods with respect to the period with the lowest count-rate (P2 in Fig.~\ref{fig_lightcurve1}).

\subsubsection{Difference spectra}

\begin{figure}[!t]
\centering{
\mbox{\hspace{-1.cm}\subfigure{\includegraphics[scale=0.54, clip=true]{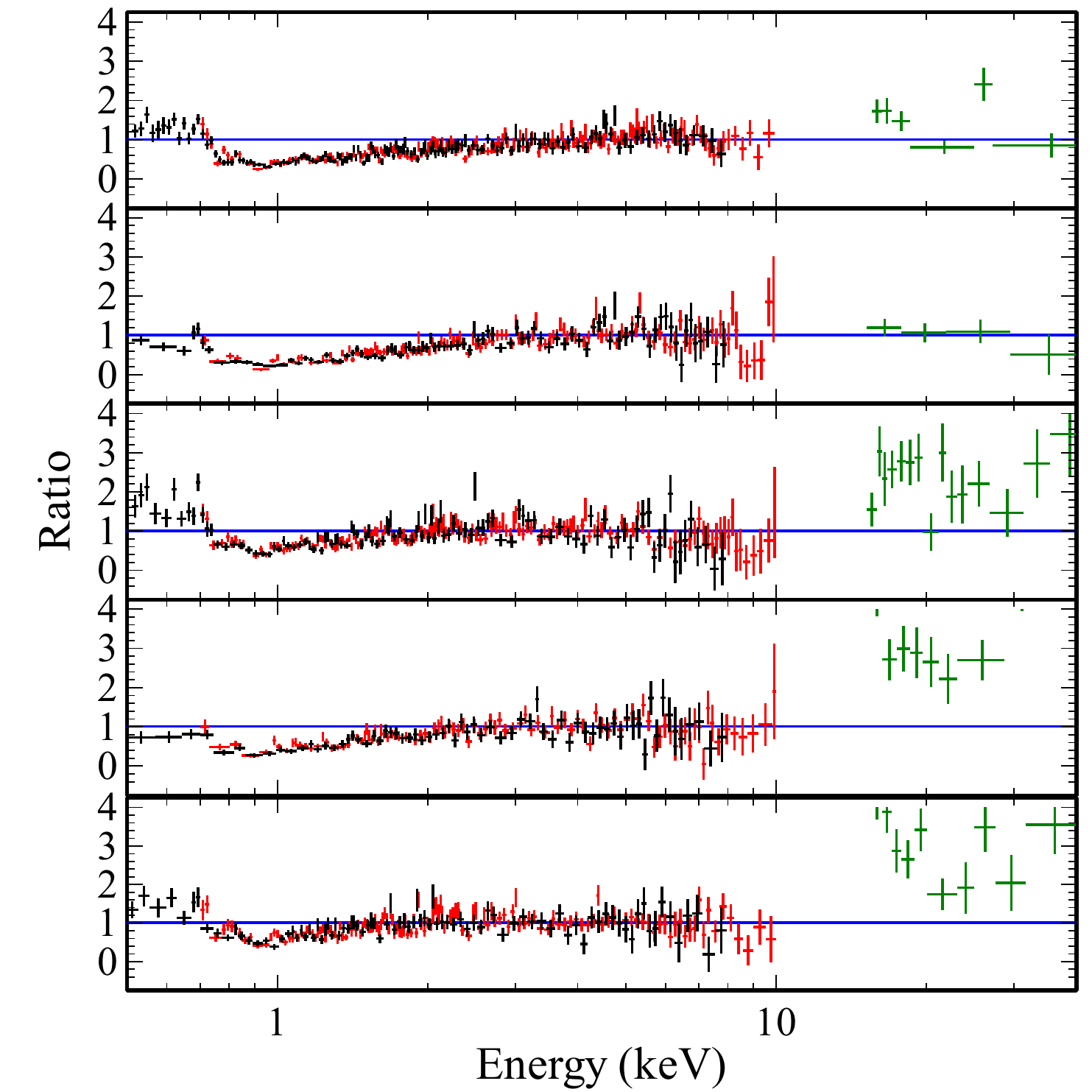}
\quad
\subfigure{\includegraphics[scale=0.54, clip=true]{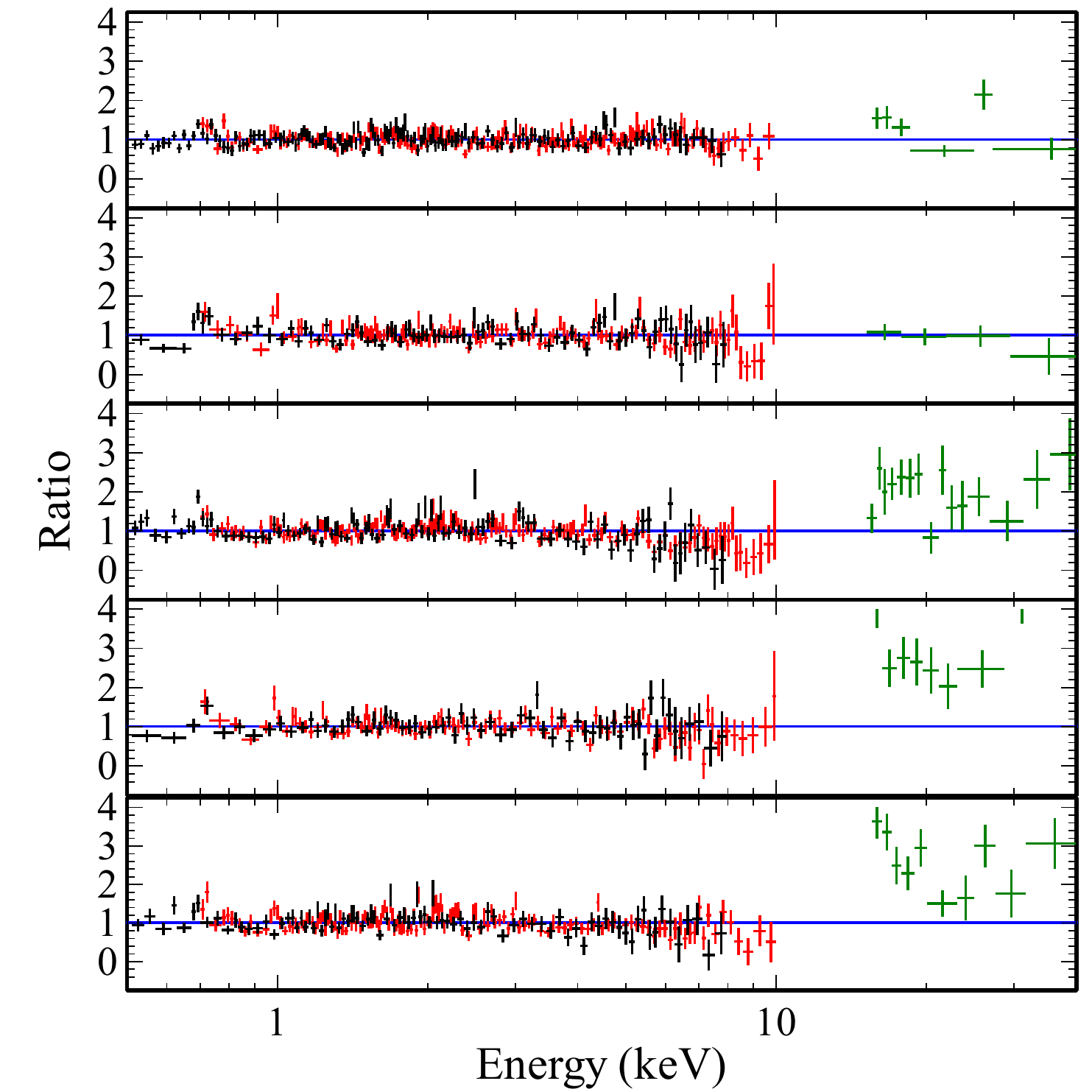} }}}}
\vspace*{-0.35cm}
\caption{ {\it Left:} Data/model ratio for the difference-spectra between (from top to bottom) P1, P3, P4, P5 and P6 minus P2. The difference spectra shown on the left are fit between 3 and 10\kev\ with a powerlaw and extended to the full energy range. {\it Right:} After the inclusion of a 3-zone warm absorber fitted between 0.5--10\kev. In both cases, the P4, P5 and P6 difference-spectra show a clear hard excess which is not an artefact of modelling the ionized absorber. BI, FI and PIN data are shown in black, red and green respectively.  }
   \label{fig_dif_time_pl}

 \end{figure}

Figure~\ref{fig_dif_time_pl} (Left) shows the data/model ratio for the five time-resolved difference-spectra fitted with an absorbed powerlaw above 3\kev. For the top two difference-spectra (P1 and P3 minus P2), a powerlaw with an index fixed $1.95$, consistent with the value found in the previous section, was satisfactory, with $\chisq/\nu = 686.0/726~(0.94)$. Allowing $\Gamma$ to vary resulted in slight improvement to the fit ($\Delta \chisq = 15.6$) and a lower value for the powerlaw index, with $\Gamma=1.80\pm0.06$. When we add the latter three difference-spectra (P4, P5 and P6 minus P2), we find that, whereas the 3-10\kev\ range is well fit with a powerlaw having an index of $\sim1.95$, the flux above 10\kev\ is severely under-predicted. This is highlighted in the bottom two panels of Fig.~\ref{fig_dif_time_pl} (Left) where $\Gamma=1.95$. Allowing $\Gamma$ to vary in the second half of the observation resulted in a statistically acceptable fit ($\chisq/\nu = 1793.7/1747 (1.03)$) however it required uncharacteristically low values for $\Gamma$ of $\approx 1.38-1.44$. Indeed, if we replace the single powerlaw with a broken-power model having a global gamma below 10\kev\ ($\Gamma_{<10{\rm keV}}$) and allow $\Gamma_{>10{\rm keV}}$ to vary between the periods, the various difference spectra, including the hard excesses, can be phenomenologically modelled ($\chisq/\nu = 1687.5/1744$) with $\Gamma_{<10{\rm keV}} = 1.88^{+0.06}_{-0.07}$ and 
$\Gamma_{>10{\rm keV}} = 1.8\pm0.2, 1.9\pm0.3, 1.07^{+0.17}_{-0.15}, 0.73^{+0.16}_{-0.13}$ and $0.87^{+0.16}_{-0.13}$ for P1, P3, P4, P5 and P6 minus P2 respectively. It is clear that, whereas the first half of the observation does not require a break in the powerlaw, the second half requires the presence of an extra hard component.

Figure~\ref{fig_dif_time_pl} (Right) shows the data/model ratio for the various difference-spectra after the inclusion of a three-zone warm absorber similar to that of the previous section (\S~\ref{flux_resolved}). The fit was obtained by modelling simultaneously all spectra in the \xis\ energy range (0.5--10\kev) together with the \pin\ data for period 1 and 3 only. The \pin\ data for periods 4, 5 and 6 emphasise the level of excess above the unbroken powerlaw continuum. The best fit model consisting of a ($\Gamma = 1.95$) powerlaw  fit to the full energy range but excluding the \pin\ data for the second half of the observation, resulted in $\chisq/\nu =  3198.7/3189~(1.00)$. Including the \pin\ data for periods 4, 5 and 6 results in a worst fit ($\chisq/\nu =  3518.5/3228~(1.09)$) with the excess above 10\kev\ being the prominent source of residuals. Allowing the powerlaw indices  to vary between the various difference-spectra did not remove the hard excess in the latter half of the observation. We show in Table~\ref{table3} the values obtained for $\Gamma$ as a function of time using this simple absorbed powerlaw model.  We emphasize that since there is still a distinct hard excess in the latter half of the observation, these values are only illustrative.

The presence of such an excess at hard energies ($> 10.0$\kev) in the difference-spectrum implies that there must be a further varying component whose contribution is greatest at these energies. Furthermore, this variation does not occur monotonically  with flux, since such hard excesses were not present in the various difference-spectra presented in \S~\ref{flux_resolved}. The fact that in all difference-spectra the energy range below $\approx10$\kev\ was well modelled with a simple powerlaw further suggests that this extra component contributes little to the variability below this energy. A possible origin for the spectral variation at high energies as a function of time could be due to change in the ionisation state of the innermost accretion disk and/or change in the contribution to the spectrum from the reflection component. The latter can be parametrised by the ratio ${\cal R}$ between the reflection and illuminating powerlaw flux. This possibility will be investigated further in the following section.

\begin{table}[!t]
\begin{center}
\caption[]{\label{table3}Summary of fits to the various time-resolved difference-spectra. }
\vspace*{0.1cm}
{%
\newcommand{\mc}[3]{\multicolumn{#1}{#2}{#3}}
\begin{center}
\begin{tabular}{l|clll}
\hline
\hline
\mc{1}{l|}{×} & \mc{4}{c}{Time-Resolved Difference-Spectra$^a$}\\\cline{1-5}
\mc{1}{c|}{Period$^b$} & Gamma Untied & \mc{1}{c}{Gamma Tied}  \\
\mc{1}{c|}{P1 - P2} & $1.95\pm0.02$          & \mc{1}{c}{$1.95\pm0.05$ } \\
\mc{1}{c|}{P3 - P2} & $1.95^{+0.02}_{-0.03}$ & \mc{1}{c}{---} \\
\mc{1}{c|}{P4 - P2} & $2.01^{+0.04}_{-0.02}$ & \mc{1}{c}{---}\\
\mc{1}{c|}{P5 - P2} & $1.68^{+0.05}_{-0.03}$ & \mc{1}{c}{---} \\
\mc{1}{c|}{P6 - P2} & $1.87^{+0.06}_{-0.02}$ & \mc{1}{c}{---} \\
\mc{1}{c|}{$\chisq/\nu$} & 3496.7/3224  & \mc{1}{c}{3518.5/3228 }\\ 
\hline
\hline
\end{tabular}
\end{center}
}%
\end{center}

{\small Notes.-$^a$All cases refer to fits performed in the full 0.5--40\kev energy range. $^b$ The difference-spectra are made from the periods highlighted in Figs.~\ref{fig_lightcurve1} and \ref{fig_image}. This model successfully accounts  for the \pin\ data during the first half of the observation but severely under-predicts the high energy flux in the difference-spectra in the latter half.}

\end{table}

\vspace*{-0.9cm}
\subsection{Fitting the individual time segments}
\label{sec:timesegments}

We have shown in the previous sections that the difference-spectra --- a rough indication of the variable components in the total observed spectrum --- can be successfully modelled using a simple  powerlaw above 3\kev\ during all flux levels, however, when the spectrum is broken down in time, this simple powerlaw does not describe the latter half of the observation. The difference-spectrum for the second half of the observation requires a further variable component which imprints itself most notably at hard energies.  In all cases, when the powerlaw continuum is extended to lower energies it reveals the strong presence of a warm absorber whose ionization and column density does not
vary with time or flux.  A self-consistent origin for both the constant hard-component as well as the varying component above 10\kev\ (other than the varying powerlaw) lies in the intrinsic shape of a reflection continuum arising from the innermost regions around a black hole for a variety of ionisations. In the following we adopt such a reflection component as the origin of both the hard-constant component and the further variability over the powerlaw component above 10\kev. We note that a reflection continuum as such is indeed usually invoked to explain the hard component in various Seyfert 1s AGN (see e.g. \citealt{Zoghbi10, waltonreis2010}).

\subsubsection{Three-component model}
\label{individual}

In order to model the individual time segments as shown in Fig.~\ref{fig_lightcurve1} we construct a model based on the primary source of the intrinsic powerlaw continuum being located in a region of strong gravity, and gravitational light bending focusing some of the emission from this component down onto the disk. The reflection component was calculated with the \reflionx\ code by \cite{reflionx}, convolved with the \kdblur\ kernel (\citealt{kdblur}) to account for the relativistic effects present close to the black hole. In Paper~1 we used the highly sophisticated variable-spin relativistic smearing model \relconv\ \citep{relconv} as the goal was to constrain the spin parameter. However, for this study of broadband X-ray variability, such a complex convolution model results in very large computational time with no extra benefit beyond that which \kdblur\ can provide. 

We assume a radial dependence for the emissivity index and ionisation parameter of the disk, such that within \rbr\ the disk is highly ionised with a variable ionisation parameter resulting in reflection which is modelled with an inner, hot, relativistically-convolved \reflionx, and beyond this radius it has a constant value of $\xi=1$\ergcmps\ up to an outer radius fixed at 400\rg\ (modelled with an outer, cold, blurred \reflionx). After several trials, we fixed the break radius at 30\rg. Changing this value to anywhere greater than approximately 6\rg\ does not change the results presented below.  To account for the narrow {\rm Fe~XXVI} emission line present in the spectra we add a Gaussian centered at 6.97\kev\ with $\sigma=10$\ev. The normalisation of this line is kept tied between each period and we find its equivalent width to vary between approximately 27\ev\ to 36\ev, in close agreement to the value found in the time average spectrum ($W_{\rm FeXXVI}=22\pm5$\ev; Paper~1). Reflection from distant, low-velocity, cold matter is modelled with \pexrav\ \citep{pexrav} together with a further narrow Fe-$\rm K\alpha$ Gaussian at 6.4\kev. When employing this model, the reflection fraction and  folding energy of the cold reflector were frozen at ${\cal R_{\rm cold}}=0.5$ and 200\kev\ respectively, similar to the values found in Paper~1. We note here that this model differs slightly from that employed in Paper~1 in that the previous paper used a superior model (\pexmon; \citealt{pexmon}) which self-consistently accounts for the Compton backscattered reflection continuum as well as the $\rm K\alpha$  and $\rm K\beta$ emission lines of iron, the Compton shoulder of the iron  $\rm K\alpha$ line, and the  $\rm K\alpha$ line of nickel. These fine details are extremely important when modelling the narrow energy range around the broad iron line and thus determining spin, however for the broadband variability we are interested in this work, the commonly used combination of \pexrav +\gaussian\ suffices. In all fits, the equivalent width of the iron  $\rm K\alpha$ line lies in the 80--100\ev\ range (compared with $W_{\rm K\alpha}=98\pm5$\ev\ and $88\pm6$\ev\ for \suzaku\ and \chandra\ data respectively; Paper~1).

Paper~1 investigated the effect of: (1) fixing the iron abundance of the distant reflector (\pexrav\ in our case) to that of the inner accretion disk and; (2) allowing these abundances to vary independently. The authors  find that a high iron abundance (Fe/solar $\approx 4$) is  statistically preferred, however little difference is found between the two scenarios. Here we fix the abundance of the  disk reflection (both inner and outer \reflionx) at 4 times solar, similar to the value used in Paper~1 but keep the distant, cold reflector fixed at solar. The Galactic neutral hydrogen column density and inner disk inclination were fixed at $9.91\times10^{20}$\pcmsq\ and 23 degrees as found in Paper~1. The powerlaw indices  of both  \reflionx\ components  are tied to that of the main continuum. The emissivity profile beyond \rbr\ is assumed to follow $\epsilon(r) \propto r^{-3}$.  We account for the effects of the three-zone warm absorber by the use of the \xstar\ grids described in \S~\ref{flux_resolved} with the values for the ionisations parameters frozen to that found in Paper~1.  Similarly to Paper~1, we also allow a fraction $f$ of the continuum to scatter/leak around the warm absorber. This value has consistently been found to be around 15 per cent. Emission from a black-body with temperature of $\approx55$\ev\ is also included in the fit.  The total model thus consists of seven  global parameters -- the inner disk radius together with the inner disk emissivity index, the powerlaw index $\Gamma$, black-body normalization, normalizations of the two narrow emission lines (Fe-$\rm K\alpha$ and {\rm Fe~XXVI})  and the fraction of scattered continuum $f$. In addition to these seven global parameters, we also  have a further seven parameters individual to each period -- column densities of the three \xstar\ grids, inner reflection ionization parameter, powerlaw normalization and the normalization of both inner  and outer reflection components. This model resulted in a good fit to all the individual periods  ($\chisq/\nu =  4494.7/3372$).

\begin{figure}[!t]
\centering
{
 
{\includegraphics[scale=0.7]{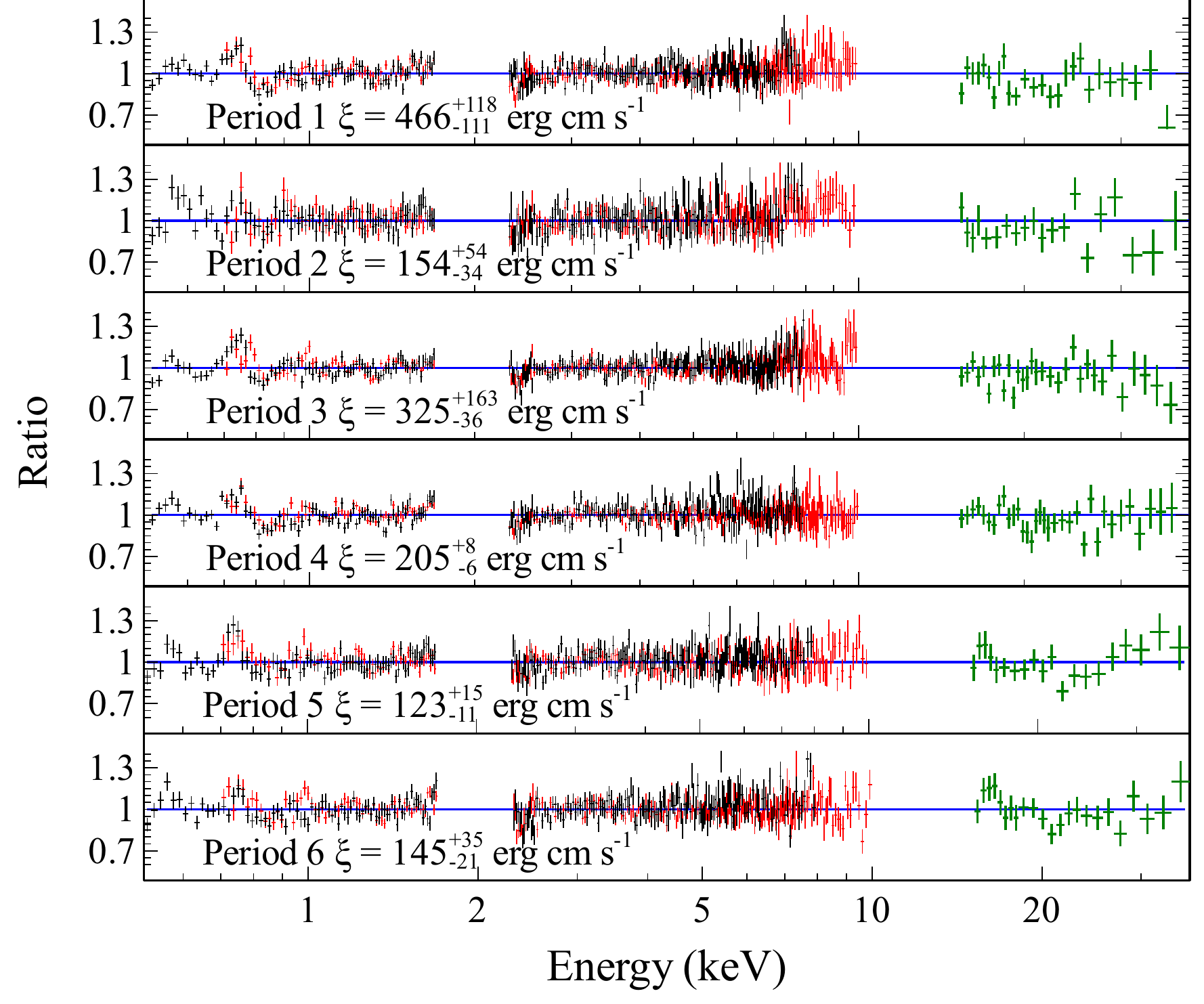}  
}}
\vspace*{-0.4cm}
\caption{Data/Model ratio for the six periods fitted simultaneously with a model consisting of a powerlaw, a cold ($\xi=1\ergcmps$) reflection component originating in the outer parts of the accretion disk and a ``hot'' reflection component with variable ionisation ($\xi$ between $\sim$100 -- 500$\ergcmps$ shown in the figure) originating from the innermost regions of the accretion disk. The continuum is absorbed by the three-zone warm absorber described previously, and the fit includes two narrow emission lines (Fe-$\rm K\alpha$ and $\rm Fe XXVI$)  as well as reflection from cold, low velocity matter as described fully in \S~\ref{individual}. See Fig.~\ref{fig_5models} for a visual depiction of the evolution of these three components during the different periods. }
\label{fig_full}
\end{figure}

\begin{table}[!t]
\begin{center}
\caption[]{\label{table4} Summary of fit to the various time-resolved spectra. }
\vspace*{-0.3cm}
\begin{center}
\begin{tabular}{lcccccc}
 \hline
 \hline
 Parameter & P1 & P2 & P3 & P4 & P5&P6\\
 \hline
\nh\ ($\pcmsq$)$^a$ & $9.91\times10^{20}$(f) & --- & --- & ---& ---& ---\\
$KT_{\rm BB}$ (\ev) & 54.6(f) & ---×× & ---×× & ---×× &--- ××&--- ××\\
Flux$_{\rm BB}$ $^b$ & $ 3.1\pm1.8$ & --- & --- & --- & ---&--- ××\\
$\Gamma$ & $1.82\pm0.01$ & --- & --- & --- & ---&--- ××\\
$q_{r \leq r_{\rm br}}$  & $4.3\pm0.4$ & ×--- & ×--- & ---× & ---×&--- ××\\
\rin (\rg) & $<1.4$ & ---× & ---× & ---× & ---×&--- ××\\
Inclination & 23(f) & ---× & ---× & ---× & ---×&--- ××\\
Fe (${\rm Z_{Fe}}$) & 4(f) & ---× & ---× & ---× & ---×&--- ××\\
$\xi$ (\ergcmps) & $466^{+44}_{-186}$& $148^{+55}_{-33}$ & $284^{+192}_{-33}$ & $205^{+8}_{-13}$ & $123^{+15}_{-10}$ &$145^{+33}_{-16} $\\
Flux$_{\rm powerlaw}$   & $37.2\pm0.7$& $29.0\pm0.6$ & $34.2\pm 0.6$ &  $31.5^{+0.7}_{-0.8}$ & $30.9^{+0.5}_{-0.6}$& $30.7^{+0.5}_{-0.6}$\\
Flux$_{\rm Inner~ ref.}$  & $7.5{+0.6}_{-0.4}$ & $2.4\pm0.6$ & $5.9^{+0.5}_{0.3}$ &$8.7\pm0.5$ & $8.9\pm0.9$ & $8.8\pm0.9$\\
Flux$_{\rm Outer~ ref.}$   & $1.9\pm0.5$ & $2.8^{+0.6}_{-0.7}$ & $1.7\pm0.5$× & $1.0^{+0.5}_{-0.4}$× & $1.3\pm0.6$&$1.6\pm0.6$\\
$\chisq/\nu$ &4497.7/3390 (1.33)& ---× & ---× & ---× & ---×&---\\
\hline
\hline

\end{tabular}
\end{center}
\end{center}
\vspace*{-0.2cm}
{\small Notes.- Fit to the individual periods highlighted in Figs.~\ref{fig_lightcurve1} and \ref{fig_image}. The six periods were fit simultaneously with a number of global, as well as period-dependent, parameters. The model and the various parameters  are described in section~3.5.1. $^a$ The hydrogen column density, disk temperature, inclination and iron abundance are frozen (f) at the values quoted in Paper~1. $^b$ The various fluxes are found between 0.001 and 1000\kev\ and are in units of $\times 10^{-11}$\ergpcmsqps. }

\end{table}

\begin{table}[!bt]
\caption[]{\label{table5}Summary of fits having the power law both tied and untied between the six periods. }
\vspace*{-0.7cm}
{%
\newcommand{\mc}[3]{\multicolumn{#1}{#2}{#3}}
\begin{center}
 \begin{tabular}{lcccccccc}
\hline
\hline
Gamma & Period 1&Period 2& Period 3& Period 4& Period 5&Period 6& {$\chisq$/d.o.f.}\\
\hline
Tied & \mc{6}{c}{$1.82\pm0.01$} &\mc{1}{c}{4497.7/3390}\\
Untied & $1.82\pm0.01$ & \mc{1}{c}{$1.82^{+0.02}_{-0.03}$} &  $1.83^{+0.02}_{-0.01}$ &$1.83^{+0.02}_{-0.01}$ & $1.82^{+0.02}_{-0.01}$& $1.83\pm0.02$&4487.4/3385\\
\hline
\end{tabular}
\end{center}
}%
\end{table}

\begin{figure}[!th]

\centering{
 \rotatebox{0}{
  \resizebox{!}{7cm} 
{\hspace*{-2.cm}\includegraphics{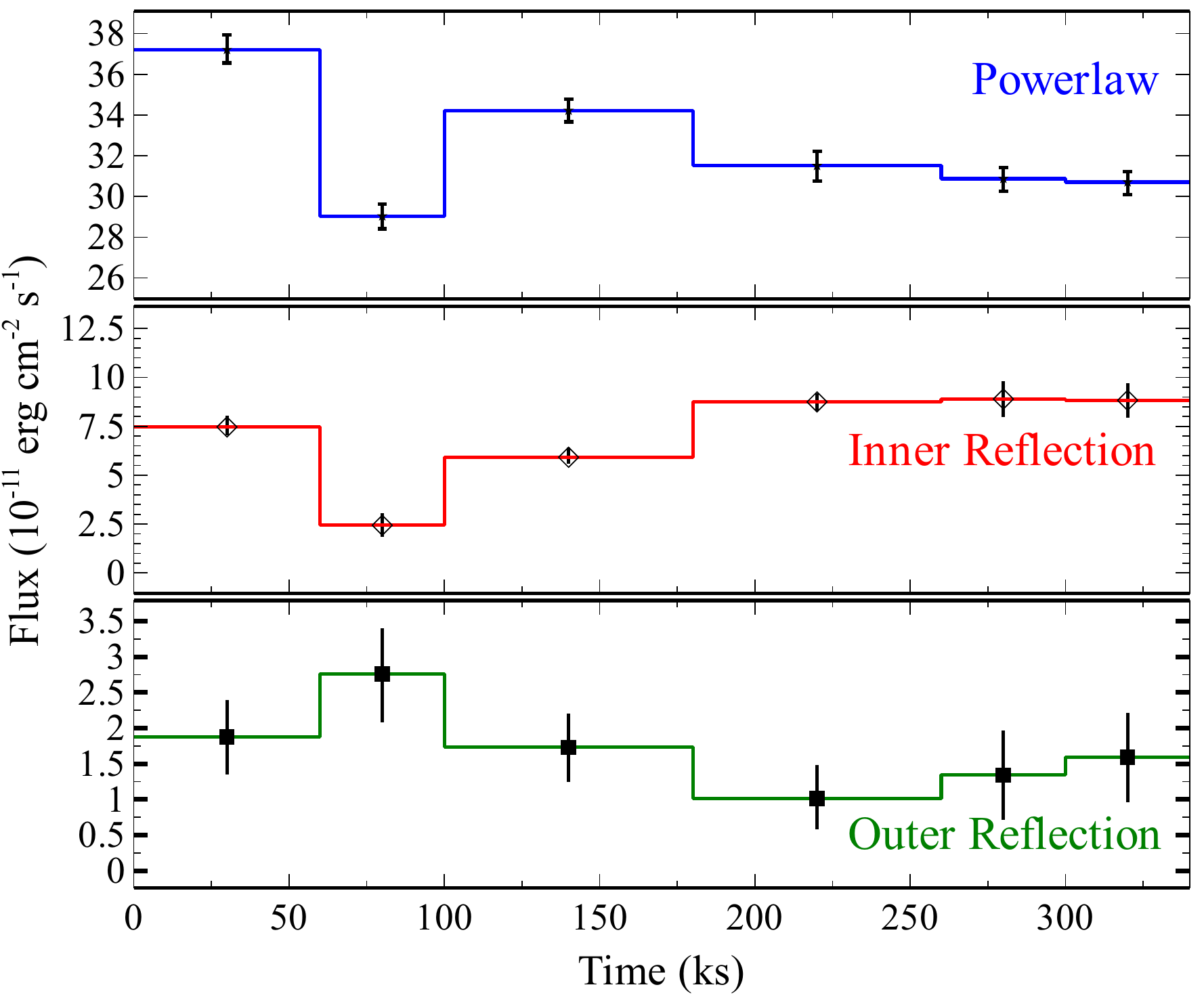}  
}}}
\centering{
 \rotatebox{0}{
  \resizebox{!}{7cm} 
{\includegraphics{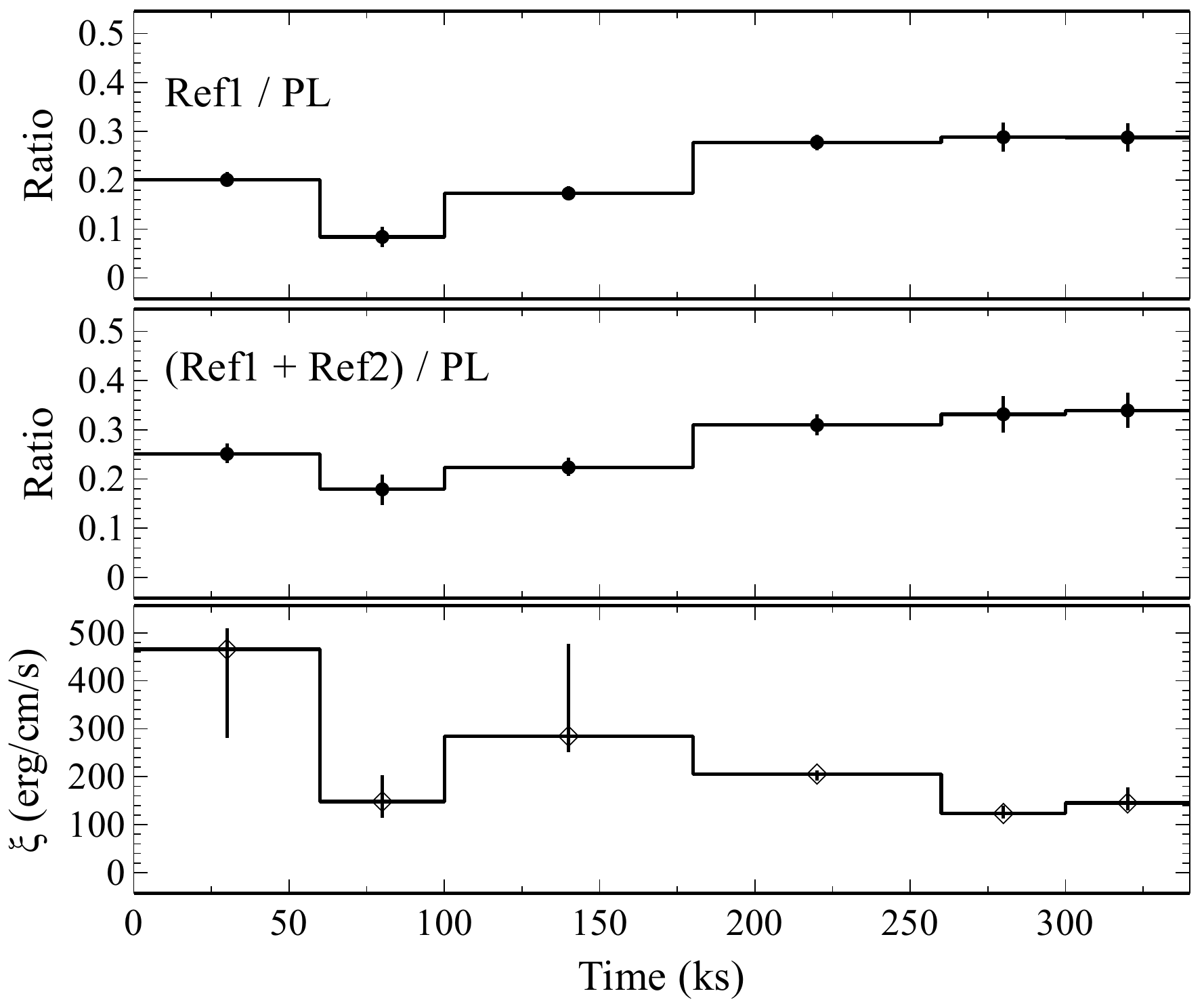}  
}}}
\vspace*{-0.3cm}
\caption{ {\it Left:} Flux evolution of the three variable components with the powerlaw, inner and outer reflection shown in blue, red and green respectively.  {\it Right:} Ratio of the inner reflection (top) and total reflection (middle) to the powerlaw flux.  The bottom panel shows the evolution of the ionisation parameter $\xi$.}
\label{fig_flux_evolution}

\vspace*{0.2cm}
\hspace*{-0.7cm}
\hbox{
{\rotatebox{270}{{\includegraphics[width=4.25cm]{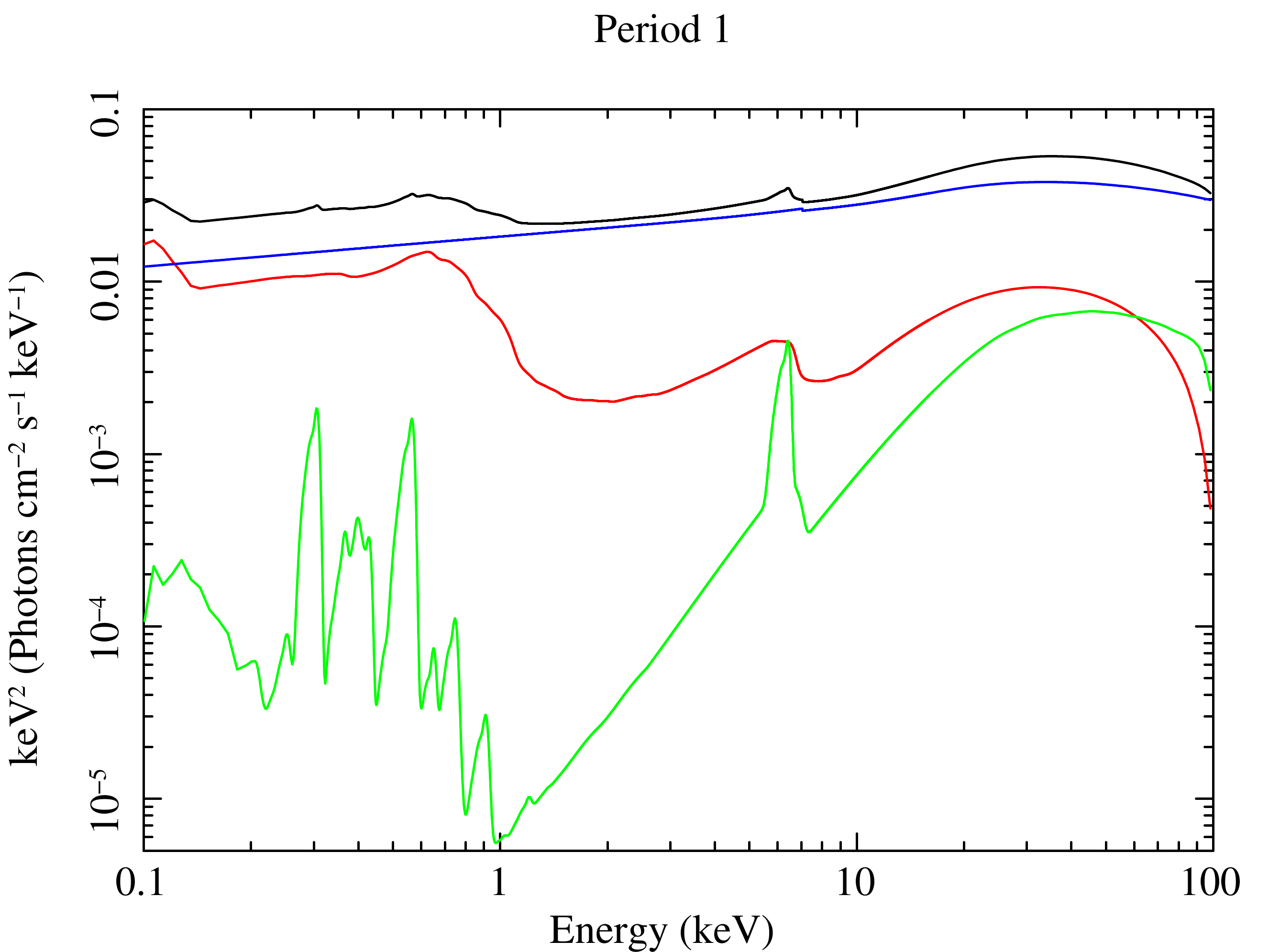}}}}
{\rotatebox{270}{{\includegraphics[width=4.25cm]{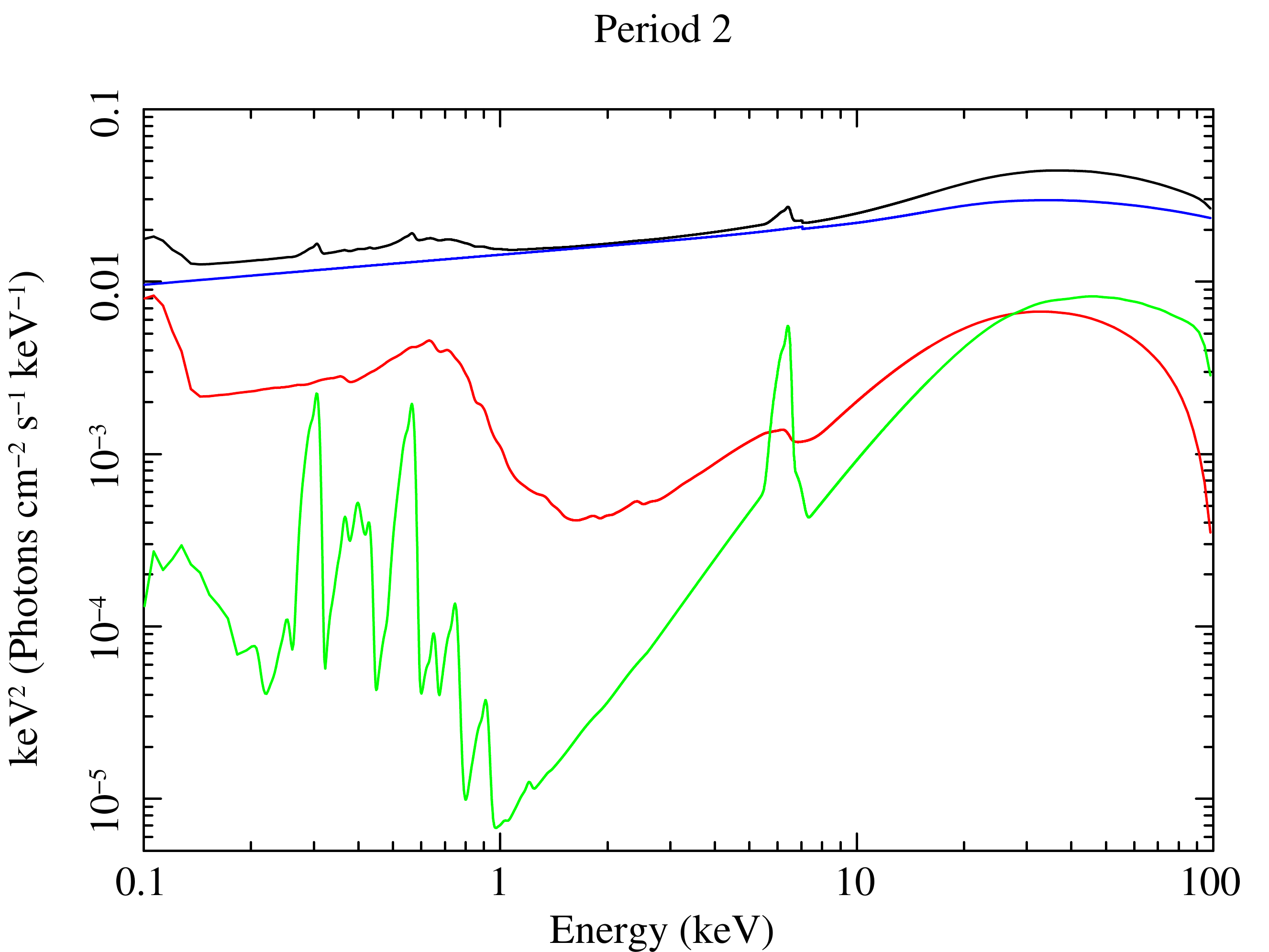}}}}
{\rotatebox{270}{{\includegraphics[width=4.25cm]{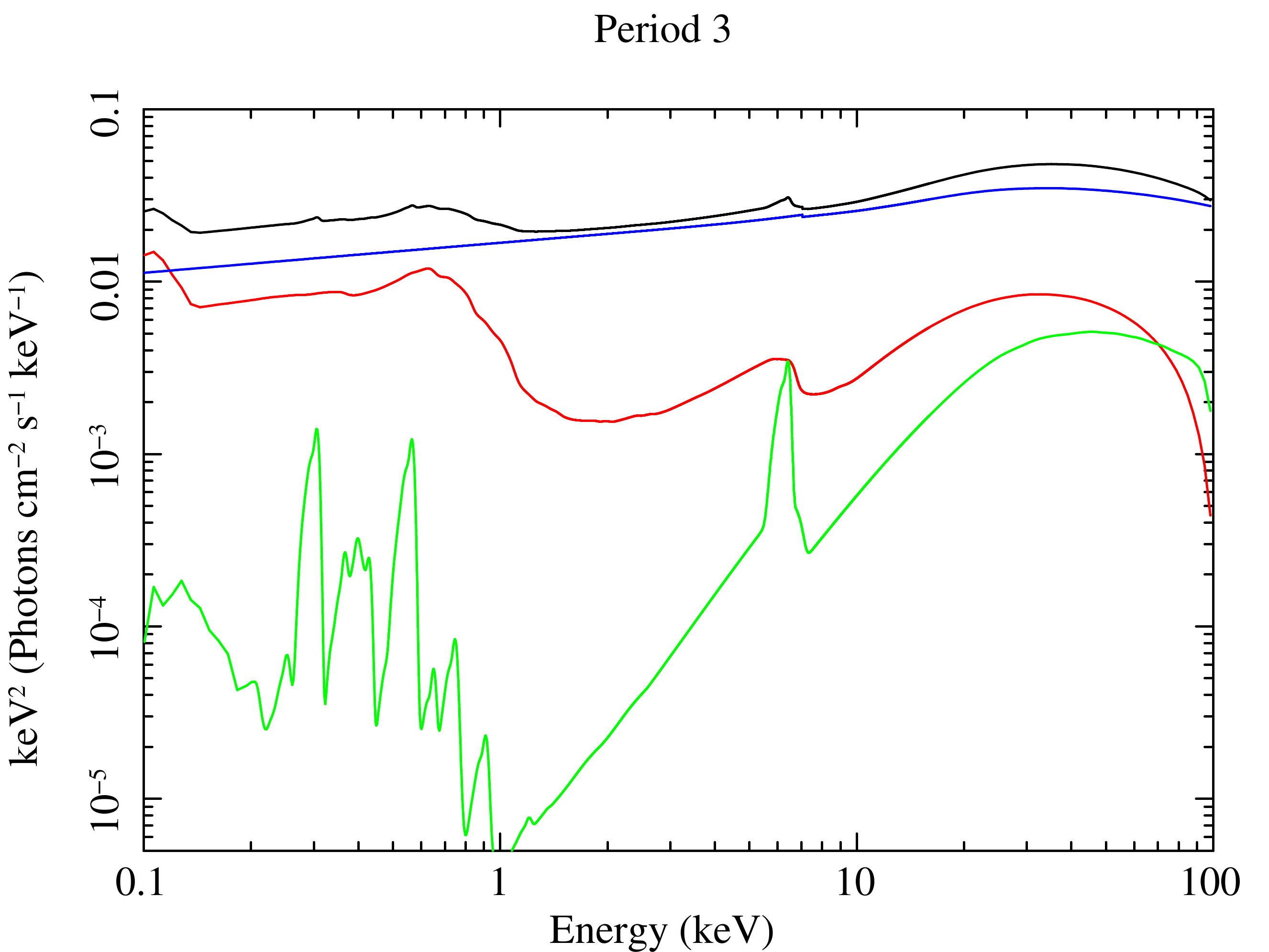}}}}}
\hspace*{-0.7cm}\hbox{
{\rotatebox{270}{{\includegraphics[width=4.25cm]{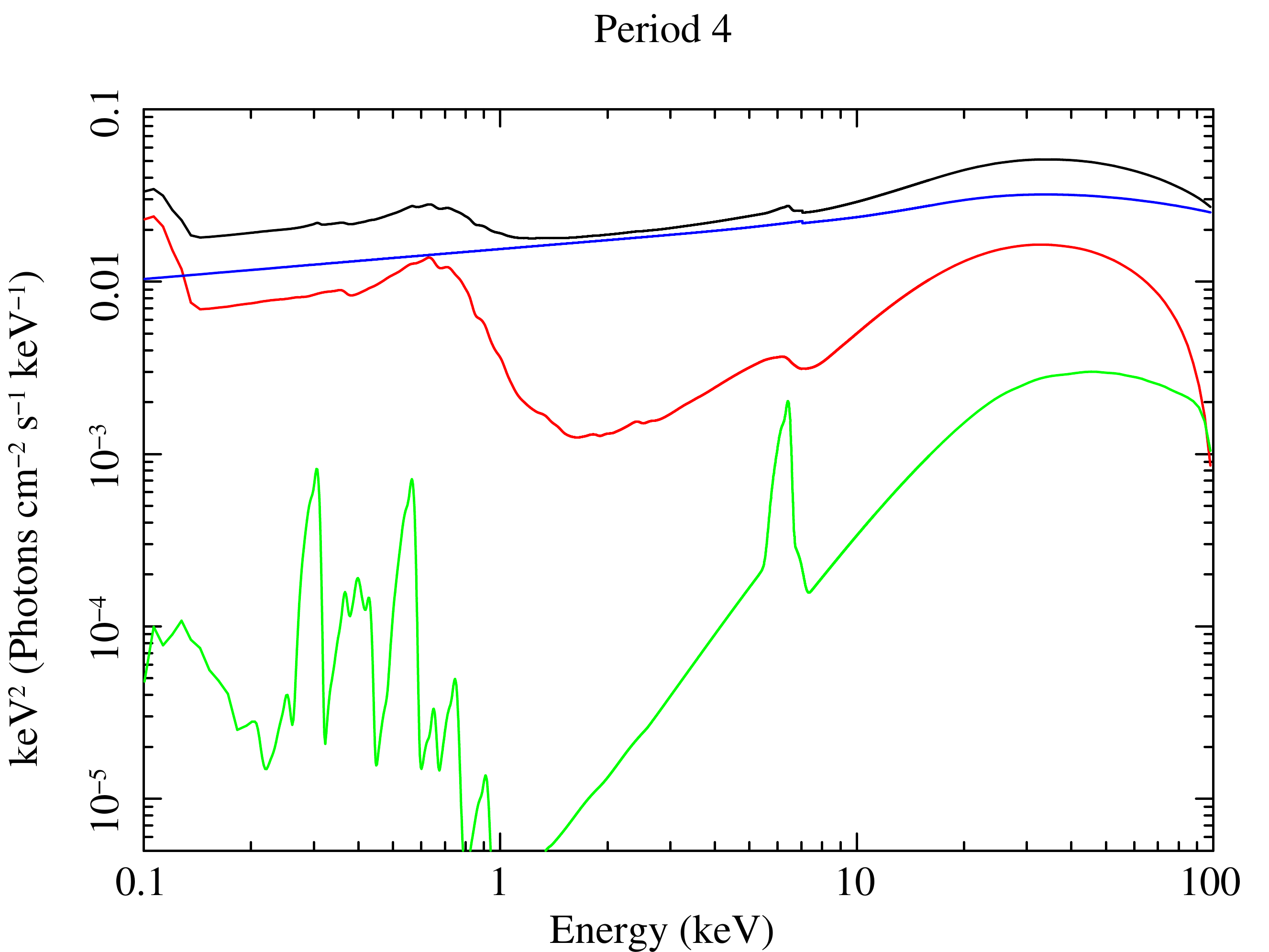}}}}
{\rotatebox{270}{{\includegraphics[width=4.25cm]{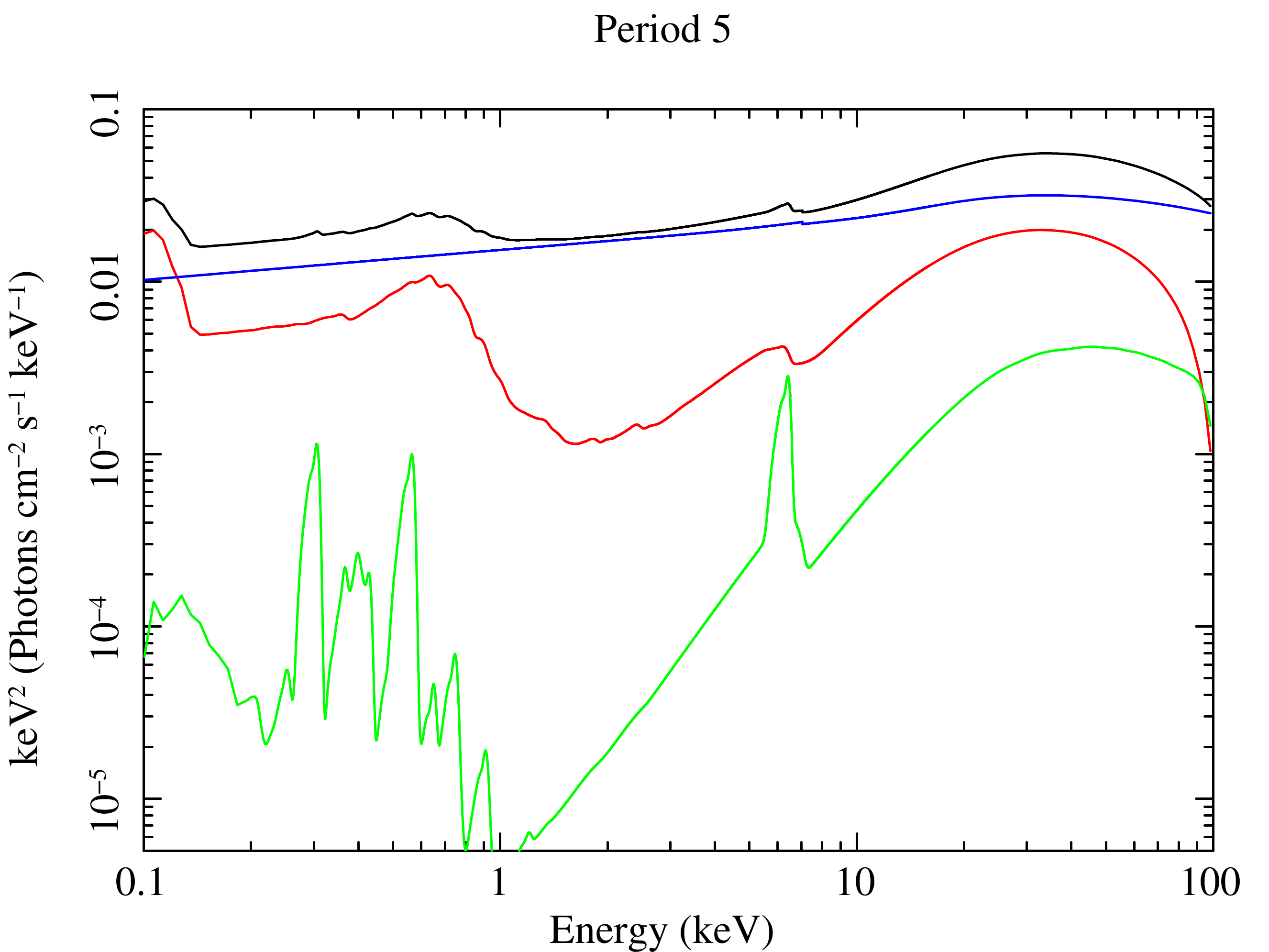}}}}
{\rotatebox{270}{{\includegraphics[width=4.25cm]{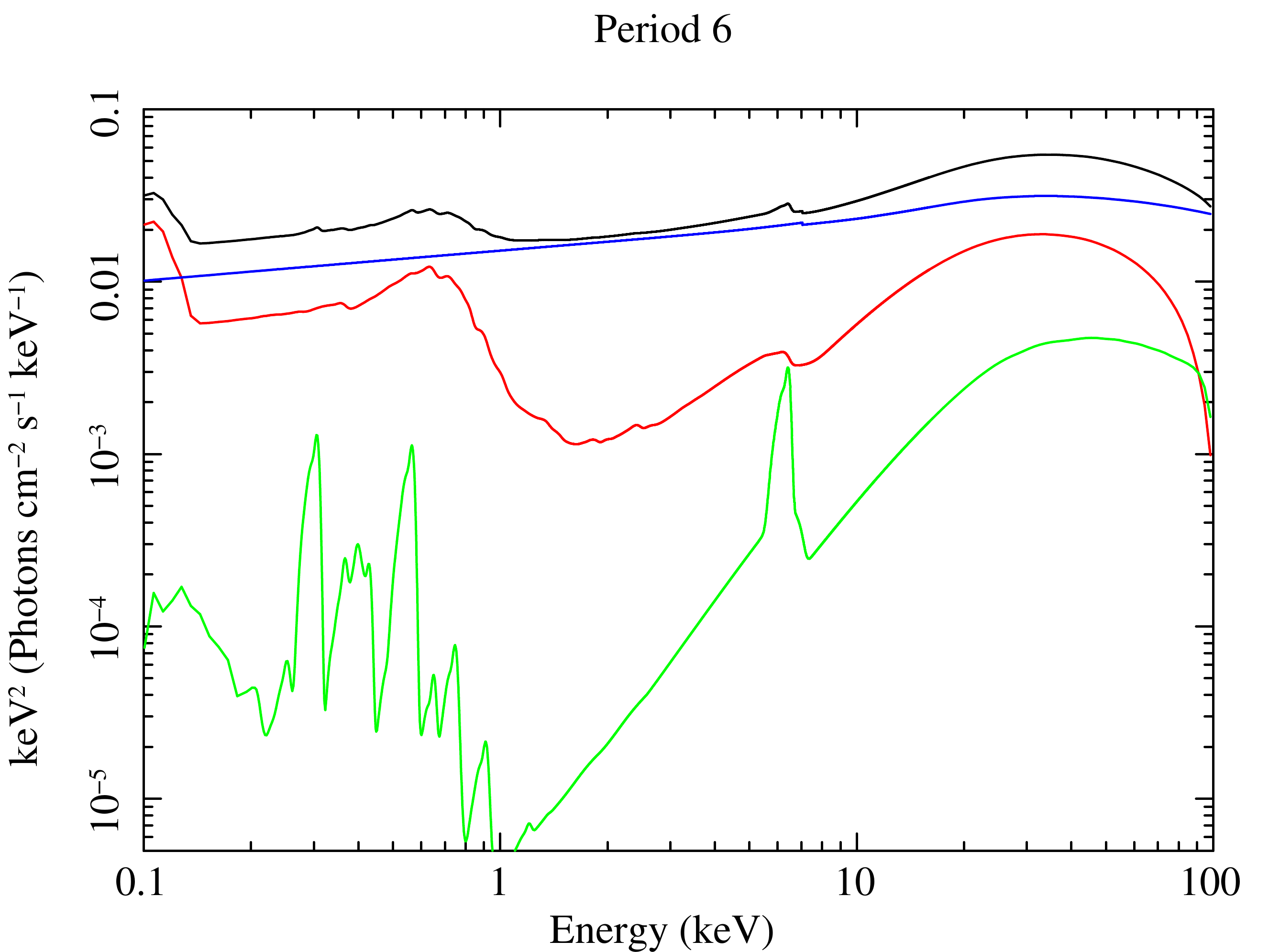}}}}}

\caption{Best-fit model for the six periods, as described in \S\ref{individual}. The total model is shown in black, with the powerlaw, inner and outer reflection components shown in blue, red and green respectively. The global blackbody, absorption and narrow emission lines were removed for display. }
\label{fig_5models}
\end{figure}

Figure~\ref{fig_full} shows the data/model ratio for all six periods and Table~\ref{table4} summarises the values found for the various parameters of interest.  In order to obtain the errors (90 per cent confidence) for the various parameters of interest in this study --- mainly the inner-disk ionisation, fluxes of the various components and the global powerlaw index --- we froze the column densities of the warm absorber to the best-fit values obtained as we have already shown that the absorption does not vary on the time scales probed in this observation. Freezing these parameters was done in order to reduce computational time and could result in a slight underestimate of the true error. Nevertheless, the global powerlaw index is found to be $\Gamma= 1.82\pm0.01$, very similar to that found for the time averaged spectrum presented in Paper~I.  We investigated the possibility of a varying powerlaw index and show in Table~\ref{table5} that gamma is perfectly consistent with the global value found above throughout the six periods. This supports the results presented in the previous section in which the difference spectra below 10\kev\ is shown to be well modelled with a simple powerlaw.

The flux evolution of the three variable components is shown in Fig.~\ref{fig_flux_evolution} (left). It is clear that as the $[10^{-3},10^{3}]$~\kev\ flux in the powerlaw continuum (blue) begins to decrease, the inner (red)  reflection also decreases. However, in the latter half of the observation the flux in the powerlaw decreases, whereas the inner-reflection flux shows a gradual increase. The reflection arising from the outer part of the disk appears to be nearly constant. In fact, an equally good fit ($\chisq/\nu=4514.6/3395 (1.33)$; i.e. $\Delta\chisq = 16.9$ for 5 d.o.f) can be achieved if we force it to be constant throughout all six periods  .   
The evolution of these components manifests itself in the observed evolution of the reflection ratio ${\cal R}$, where we see in the right panels (top and middle) that the total contribution to the spectrum from the reflection component reaches a maximum in the second half of the observation and has a clear minimum during Period 2. We also show in the the bottom-right panel the evolution of the inner ionisation parameter  ($\xi = L_{int}/nR^2$). Assuming the intrinsic luminosity ($L_{int}$) of the powerlaw continuum (as seen by the accretion disk) as well as the hydrogen number density of the disk ($n$) remains constant then the maximum change in ionisation observed ($\xi_{high}/\xi_{low}\approx 4$) implies an increase in the height of the corona (above the disk) by a factor of $\sim2$ (assuming Newtonian geometry). Of course it is likely that there could be some contribution from both intrinsic variations in the continuum luminosity, and possibly changes in the hydrogen number density of the disk as well. In Figure~\ref{fig_5models} we visually depict the evolution of the three components during the course of the observation. The greater contribution from the inner reflection component (red) to the hard energy band during the last three periods in comparison to the first three is clearly seen (the near constancy of the reflection component arising from the outer parts of the accretion disk -- shown in green -- can be used as a reference point to highlight the level of change in the inner reflection flux). The extra flux at hard energies coming from the increase of the inner reflection component in comparison to the powerlaw continuum directly explains the presence of a hard excess in the difference-spectra presented in \S\ref{time_resolved} for the second half of the observation  in comparison to Period~2.

%\section{Summary of observational results}

%\item The ionisation  parameters of the inner reflection component was found to change by a factor of five between the first and last period (Fig.~\ref{fig_flux_evolution}). Such a change was shown to be (possibly) due to an increase by a factor of $\sim2$ in the height of the corona above the accretion disk.

% \item The positive correlation between powerlaw and reflection flux in the first half of the observation  (Fig.~\ref{fig_flux_evolution}) suggest that the corona was  between $\sim1-4$\rg\ above the disk (Regime~1 in the light-bending model of \citealt{Miniu04}). The increase in the powerlaw flux and the subsequent decrease in reflection flux suggest that the corona moved to a height of $\sim10$\rg\ above the disk (Regime~2) towards the second half of the observation. 

\vspace*{-0.5cm}
\section{Discussion and Conclusions}
\label{discussion}

The main results concerning the spectral variability of \n\ as determined from the long \suzaku\ observation can be summarized as follows:

\begin{enumerate}
\item The spectral variability of \n\ can be described by the superposition of a soft-variable component together with a quasi-constant hard component. The shape of this quasi-constant component (Fig.~\ref{fig_flux_flux3}) is hard in the 3--10\kev\ range as found for several other AGN, and flat both below 1\kev\ and above 10\kev. These features are qualitatively similar to that expected from reflection arising from the inner parts of an accretion disk.

\item  A simple powerlaw provides a good fit to all the difference-spectra between  3--10\kev\ suggesting that in this band the variability is dominated by variation in the normalization of the powerlaw-like Comptonized continuum.

\item Extending the powerlaw to lower energies reveals the strong  presence of a warm absorber (Figs.~\ref{fig_ratio2pl},\ref{fig_dif_flux_pl} and \ref{fig_dif_time_pl}). The warm absorber does not vary with time or flux (see e.g. Fig.\ref{fig_high_low_flux_ratio2pl}), which is consistent with previous studies of this source.

\item The presence of complex spectral evolution is clearly shown in Fig.~\ref{fig_image}. The evolution of the softness ratio allows for the clear distinction of at least six ``spectral periods'' (Fig.~\ref{fig_lightcurve1}). 

\item Time-resolved difference-spectra show the presence of an excess flux to a powerlaw continuum above 10\kev\ in the latter half of the observation (Fig.~\ref{fig_dif_time_pl}). This excess is not seen in the flux-resolved difference-spectra (Fig.~\ref{fig_dif_flux_pl}) suggesting that the further variable component above 10\kev\ varies with time but not monotonically with flux. 

\item Changes in the reflection fraction and possibly ionisation state of the inner accretion disk could result in a reflection component having a quasi-constant level below 10\kev\ but different contributions at higher energies . A model consisting of such a reflection component together with a powerlaw and the associated Compton backscattered emission from cold material, provides an excellent description of the continuum for the six distinct spectral periods seen in \n\ (Fig.~\ref{fig_full}). 
\end{enumerate}

Our time-domain spectral analysis presented in Section~\ref{sec:timesegments} presents a complex picture.   To organize this discussion, we choose to focus on three primary observables, the observed power-law flux ($PL$), the ratio of the total disk reflection to the powerlaw flux ${\cal R}=(Ref1+Ref2)/PL$, and the ionization state of the inner accretion disk ($\xi$).   Thus, we have a 3-dimensional parameter space $(PL, {\cal R}, \xi)$ that, broadly, describes the X-ray flux and spectral state of the central engine.

What are the theoretical expectations for the evolution of the source in this parameter space?  We suppose for now that the observed continuum X-rays are emitted from a single structure (``the corona"), and that the observed reflection spectrum results from irradiation of the disk by that same structure.  Within this picture, we can imagine at least three distinct causes for flux- and spectral-variability.     Firstly, suppose that the X-ray emitting corona increases its luminosity while all other physical properties of the system (coronal geometry, coronal beaming, and disk density) remain constant.   Clearly both the observed flux and the ionization parameter of the inner disk would increase in proportion, i.e., we would expect $\xi\propto PL$.   Since the geometry is unchanged, however, we would expect ${\cal R}$ to remain constant.   The second type of variability corresponds to a change in the location, geometry, size or beaming pattern of the corona while its luminosity and the disk properties remain unchanged or changes in the geometry of the disk in a manner that some discrete Compton-thick clump of gas could move into the irradiation zone thus changing the (anisotropic) scattered/reflected emission  directed into the line of sight. In the case that these changes lead to increased irradiation of the disk at the expense of flux reaching the observer (through, for example, increased gravitational light bending if the source moves closer to the black hole), {\it PL} would decrease while both $\xi$ and ${\cal R}$ would increase in rough proportion.  The final type of change that we consider are density changes in the surface layers of the accretion disk (with all other aspects of the system remaining unchanged).   Such behaviour may result from viscous time-scale evolution{\footnote{For NGC~3783 the dynamical time-scale ($\Omega^{-1}$) at 6\rg\ is approximately 350\s. Assuming $T_{\rm viscous} \approx \alpha^{-1}(H/R)^{-2}\Omega^{-1}$ and $\alpha = (H/R) = 0.1$ we have  $T_{\rm viscous}\sim 350$\ks, similar to the duration of the observation. 
}} following some dramatic accretion or ejection event.  In this type of change, $\xi$ would respond inversely to density changes, while ${\cal R}$ and $PL$ would remain unchanged.

\begin{figure}[t]
\hspace*{-0.5cm}\hbox{
{\rotatebox{0}{{\includegraphics[width=5.4cm]{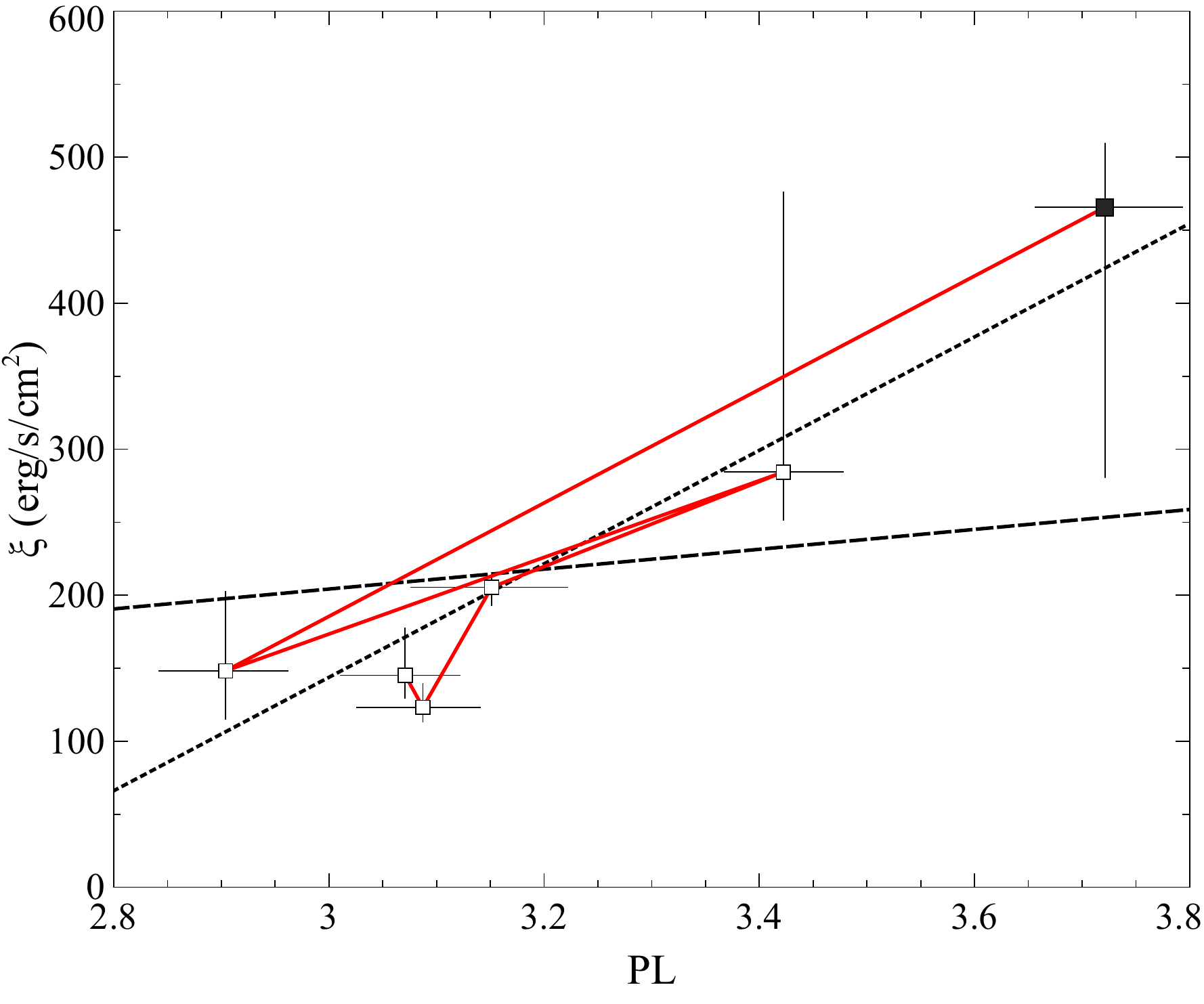}}}}
{\rotatebox{0}{{\includegraphics[width=5.4cm]{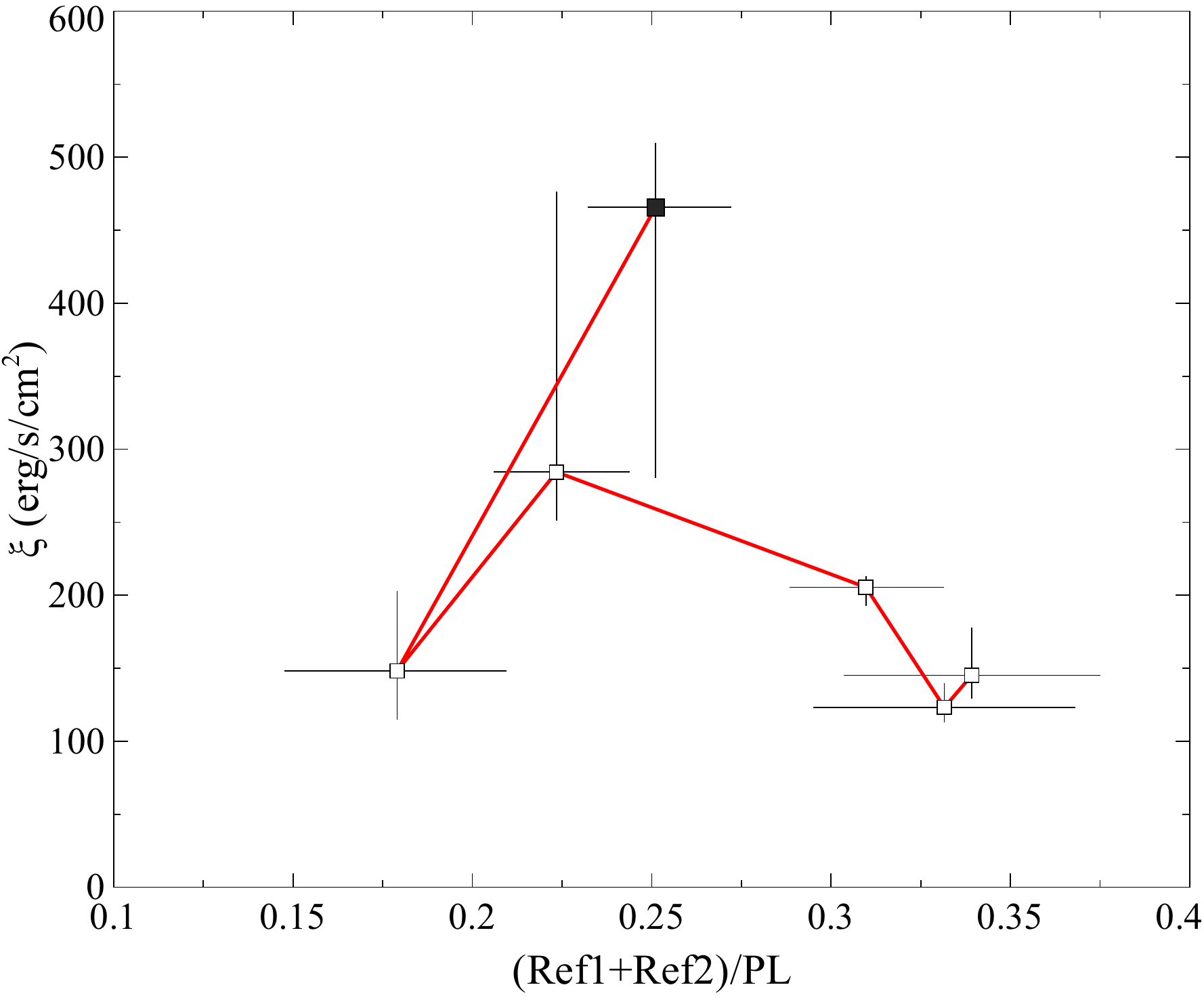}}}}
{\rotatebox{0}{{\includegraphics[width=5.4cm]{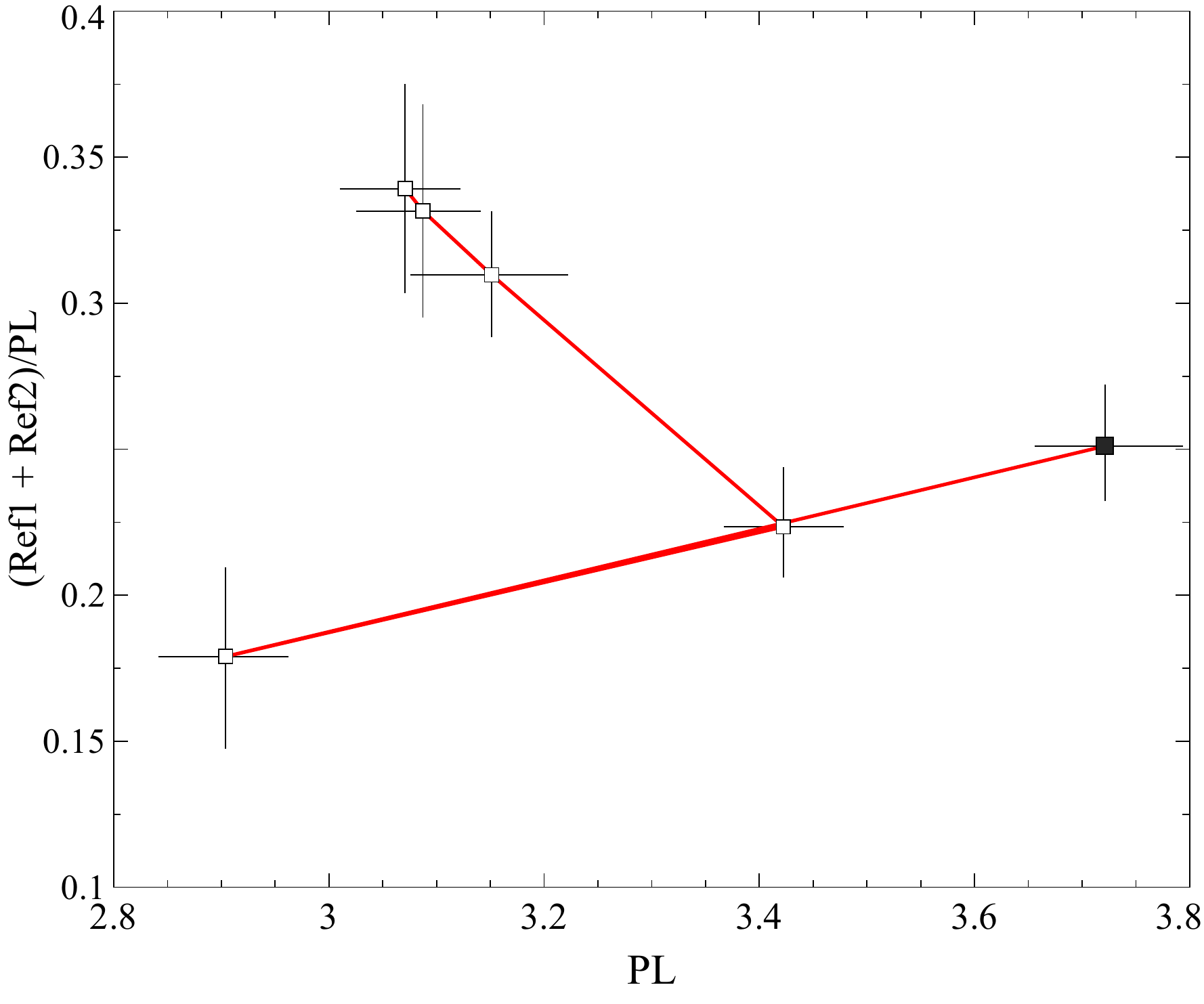}}}}
}
\vspace*{-0.5cm}
\caption{Variability of NGC~3783 in the parameter space defined by the observed power-law flux $PL$ (in units of $\times 10^{-10} \ergpcmsqps$, ionization parameter of the inner disk $\xi$ (in units of $\ergcmps$), and the ratio of the total reflected flux to the observed power-law ${\cal R}=(Ref1+Ref2)/PL$.   Lines connect data points from consecutive time segments; the first time-segment has a filled marker.   In the {\it left panel}, the dashed line shows a representative line with $\xi\propto PL$ expected from pure luminosity evolution of the X-ray corona; clearly this is a poor match to the data.  The dotted line shows the best-fitting relationship $\xi\propto (PL-PL_0)$ where $PL_0$ is some constant flux component that has little or no associated reflection.}
\label{fig:xi_pl_ref}
\end{figure}

Figure~\ref{fig:xi_pl_ref} shows the evolution of NGC~3783 projected onto the $(PL, \xi)$, $(PL,{\cal R})$, and $({\cal R}, \xi)$ planes.   Clearly, none of the three simple scenarios presented above can describe the complex behaviour that we see --- the variability must  be due to a combination of physical changes.   The fact that ${\cal R}$ changes demonstrates that there must be structural and/or geometrical changes in the X-ray emitting corona and/or accretion disk.   The most dramatic of these changes are in the final three data segments, although there is some weaker (and possibly flux-correlated) variability of ${\cal R}$ even between the first three data segments (Fig.~\ref{fig:xi_pl_ref}, right panel).

It is interesting to note that $\xi$ and $PL$  do indeed seem correlated.  However, we can only recover the linear proportionality between $\xi$ and $PL$ (expected for ``pure" coronal luminosity changes) if we assume that there is a constant flux component (corresponding to $PL=2.63\times 10^{-10}\ergpcmsqps$) that has no or very weak associated reflection spectrum, e.g., highly beamed emission from a steady jet.    In this picture, all of the observed flux variability is due to the corona which also is the principal source irradiating the disk.   In addition to reproducing the observed $\xi$-$PL$ correlations, it also explains the weak correlation of ${\cal R}$ with $PL$ during the first three time segments --- as $PL$ increases, both its contribution and the contribution of its associated reflection spectrum increase relative to the constant component. The dramatic increase in ${\cal R}$ at the end of the dataset still requires structural changes, however.

Like any data, the diagnostic power of these spectra are necessarily
limited.  However, a relatively simple and physically-motivated analysis of
the spectrum and source variability shows that reflection from a disk with
corresponding changes in ionization provides an excellent description of
the data. This discussion illustrates the power and utility of long X-ray datasets for examining AGN spectral variability.  Such studies have the potential to yield new insights into the complex interplay between various physical processes in the cores of active galaxies.

RCR is supported by NASA through the Einstein Fellowship Program, grant number PF1-120087. We are extremely grateful to our NASA and JAXA colleagues in the {\it Suzaku} project for enabling these Key Project data to be collected.   This work was supported by NASA under the {\it Suzaku} Guest Observer grants NNX09AV43G and NNX10AR31G.

\bibliographystyle{apj}

\bibliography{ref}

\begin{thebibliography}{35}
\expandafter\ifx\csname natexlab\endcsname\relax\def\natexlab#1{#1}\fi

\bibitem[{{Blustin} {et~al.}(2002){Blustin}, {Branduardi-Raymont}, {Behar},
  {Kaastra}, {Kahn}, {Page}, {Sako}, \& {Steenbrugge}}]{Blustin02}
{Blustin}, A.~J., {Branduardi-Raymont}, G., {Behar}, E., {Kaastra}, J.~S.,
  {Kahn}, S.~M., {Page}, M.~J., {Sako}, M., \& {Steenbrugge}, K.~C. 2002, \aap,
  392, 453

\bibitem[{{Boldt}(1987)}]{Boldt87}
{Boldt}, E. 1987, in IAU Symposium, Vol. 124, Observational Cosmology, ed.
  {A.~Hewitt, G.~Burbidge, \& L.~Z.~Fang}, 611--615

\bibitem[{{Brenneman} {et~al.}(2011){Brenneman}, {Reynolds}, {Nowak}, {Reis},
  {Trippe}, {Fabian}, {Iwasawa}, {Lee}, {Miller}, {Mushotzky}, {Nandra}, \&
  {Volonteri}}]{Brenneman20113783}
{Brenneman}, L.~W., {Reynolds}, C.~S., {Nowak}, M.~A., {Reis}, R.~C., {Trippe},
  M., {Fabian}, A.~C., {Iwasawa}, K., {Lee}, J.~C., {Miller}, J.~M.,
  {Mushotzky}, R.~F., {Nandra}, K., \& {Volonteri}, M. 2011, ArXiv e-prints

\bibitem[{{Dauser} {et~al.}(2010){Dauser}, {Wilms}, {Reynolds}, \&
  {Brenneman}}]{relconv}
{Dauser}, T., {Wilms}, J., {Reynolds}, C.~S., \& {Brenneman}, L.~W. 2010,
  \mnras, 409, 1534

\bibitem[{{Fabian} \& {Vaughan}(2003)}]{Fabianvaughan03}
{Fabian}, A.~C. \& {Vaughan}, S. 2003, \mnras, 340, L28

\bibitem[{{Fabian} {et~al.}(2009){Fabian}, {Zoghbi}, {Ross}, {Uttley}, {Gallo},
  {Brandt}, {Blustin}, {Boller}, {Caballero-Garcia}, {Larsson}, {Miller},
  {Miniutti}, {Ponti}, {Reis}, {Reynolds}, {Tanaka}, \& {Young}}]{FabZog09}
{Fabian}, A.~C., {Zoghbi}, A., {Ross}, R.~R., {Uttley}, P., {Gallo}, L.~C.,
  {Brandt}, W.~N., {Blustin}, A.~J., {Boller}, T., {Caballero-Garcia}, M.~D.,
  {Larsson}, J., {Miller}, J.~M., {Miniutti}, G., {Ponti}, G., {Reis}, R.~C.,
  {Reynolds}, C.~S., {Tanaka}, Y., \& {Young}, A.~J. 2009, \nat, 459, 540

\bibitem[{{Green} {et~al.}(1993){Green}, {McHardy}, \& {Lehto}}]{Greenetal1993}
{Green}, A.~R., {McHardy}, I.~M., \& {Lehto}, H.~J. 1993, \mnras, 265, 664

\bibitem[{{Halpern}(1984)}]{Halpern1984}
{Halpern}, J.~P. 1984, \apj, 281, 90

\bibitem[{{Kaspi} {et~al.}(2002){Kaspi}, {Brandt}, {George}, {Netzer},
  {Crenshaw}, {Gabel}, {Hamann}, {Kaiser}, {Koratkar}, {Kraemer}, {Kriss},
  {Mathur}, {Mushotzky}, {Nandra}, {Peterson}, {Shields}, {Turner}, \&
  {Zheng}}]{Kaspi02}
{Kaspi}, S., {Brandt}, W.~N., {George}, I.~M., {Netzer}, H., {Crenshaw}, D.~M.,
  {Gabel}, J.~R., {Hamann}, F.~W., {Kaiser}, M.~E., {Koratkar}, A., {Kraemer},
  S.~B., {Kriss}, G.~A., {Mathur}, S., {Mushotzky}, R.~F., {Nandra}, K.,
  {Peterson}, B.~M., {Shields}, J.~C., {Turner}, T.~J., \& {Zheng}, W. 2002,
  \apj, 574, 643

\bibitem[{{Koyama} {et~al.}(2007){Koyama}, {Tsunemi}, {Dotani}, {Bautz},
  {Hayashida}, {Tsuru}, {Matsumoto}, {Ogawara}, {Ricker}, {Doty}, {Kissel},
  {Foster}, {Nakajima}, {Yamaguchi}, {Mori}, {Sakano}, {Hamaguchi},
  {Nishiuchi}, {Miyata}, {Torii}, {Namiki}, {Katsuda}, {Matsuura}, {Miyauchi},
  {Anabuki}, {Tawa}, {Ozaki}, {Murakami}, {Maeda}, {Ichikawa}, {Prigozhin},
  {Boughan}, {Lamarr}, {Miller}, {Burke}, {Gregory}, {Pillsbury}, {Bamba},
  {Hiraga}, {Senda}, {Katayama}, {Kitamoto}, {Tsujimoto}, {Kohmura}, {Tsuboi},
  \& {Awaki}}]{SUZ_XIS}
{Koyama}, K., {Tsunemi}, H., {Dotani}, T., {Bautz}, M.~W., {Hayashida}, K.,
  {Tsuru}, T.~G., {Matsumoto}, H., {Ogawara}, Y., {Ricker}, G.~R., {Doty}, J.,
  {Kissel}, S.~E., {Foster}, R., {Nakajima}, H., {Yamaguchi}, H., {Mori}, H.,
  {Sakano}, M., {Hamaguchi}, K., {Nishiuchi}, M., {Miyata}, E., {Torii}, K.,
  {Namiki}, M., {Katsuda}, S., {Matsuura}, D., {Miyauchi}, T., {Anabuki}, N.,
  {Tawa}, N., {Ozaki}, M., {Murakami}, H., {Maeda}, Y., {Ichikawa}, Y.,
  {Prigozhin}, G.~Y., {Boughan}, E.~A., {Lamarr}, B., {Miller}, E.~D., {Burke},
  B.~E., {Gregory}, J.~A., {Pillsbury}, A., {Bamba}, A., {Hiraga}, J.~S.,
  {Senda}, A., {Katayama}, H., {Kitamoto}, S., {Tsujimoto}, M., {Kohmura}, T.,
  {Tsuboi}, Y., \& {Awaki}, H. 2007, \pasj, 59, 23

\bibitem[{{Krongold} {et~al.}(2003){Krongold}, {Nicastro}, {Brickhouse},
  {Elvis}, {Liedahl}, \& {Mathur}}]{Krongold2003}
{Krongold}, Y., {Nicastro}, F., {Brickhouse}, N.~S., {Elvis}, M., {Liedahl},
  D.~A., \& {Mathur}, S. 2003, \apj, 597, 832

\bibitem[{{Krongold} {et~al.}(2005){Krongold}, {Nicastro}, {Brickhouse},
  {Elvis}, \& {Mathur}}]{Krongold2005}
{Krongold}, Y., {Nicastro}, F., {Brickhouse}, N.~S., {Elvis}, M., \& {Mathur},
  S. 2005, \apj, 622, 842

\bibitem[{{Laor}(1991)}]{kdblur}
{Laor}, A. 1991, \apj, 376, 90

\bibitem[{{Magdziarz} \& {Zdziarski}(1995)}]{pexrav}
{Magdziarz}, P. \& {Zdziarski}, A.~A. 1995, \mnras, 273, 837

\bibitem[{{McHardy}(1989)}]{McHardy1989}
{McHardy}, I.~M. 1989, in ESA Special Publication, Vol. 296, Two Topics in
  X-Ray Astronomy, Volume 1: X Ray Binaries. Volume 2: AGN and the X Ray
  Background, ed. {J.~Hunt \& B.~Battrick}, 1111--1124

\bibitem[{{Miller}(2007)}]{miller07review}
{Miller}, J.~M. 2007, \araa, 45, 441

\bibitem[{{Miniutti} {et~al.}(2007){Miniutti}, {Fabian}, {Anabuki}, {Crummy},
  {Fukazawa}, {Gallo}, {Haba}, {Hayashida}, {Holt}, {Kunieda}, {Larsson},
  {Markowitz}, {Matsumoto}, {Ohno}, {Reeves}, {Takahashi}, {Tanaka},
  {Terashima}, {Torii}, {Ueda}, {Ushio}, {Watanabe}, {Yamauchi}, \&
  {Yaqoob}}]{Miniutti07}
{Miniutti}, G., {Fabian}, A.~C., {Anabuki}, N., {Crummy}, J., {Fukazawa}, Y.,
  {Gallo}, L., {Haba}, Y., {Hayashida}, K., {Holt}, S., {Kunieda}, H.,
  {Larsson}, J., {Markowitz}, A., {Matsumoto}, C., {Ohno}, M., {Reeves}, J.~N.,
  {Takahashi}, T., {Tanaka}, Y., {Terashima}, Y., {Torii}, K., {Ueda}, Y.,
  {Ushio}, M., {Watanabe}, S., {Yamauchi}, M., \& {Yaqoob}, T. 2007, \pasj, 59,
  315

\bibitem[{{Mitsuda} {et~al.}(2007){Mitsuda}, {Bautz}, {Inoue}, {Kelley},
  {Koyama}, {Kunieda}, {Makishima}, {Ogawara}, {Petre}, {Takahashi}, {Tsunemi},
  {White}, {Anabuki}, {Angelini}, {Arnaud}, {Awaki}, {Bamba}, {Boyce}, {Brown},
  {Chan}, {Cottam}, {Dotani}, {Doty}, {Ebisawa}, {Ezoe}, {Fabian}, {Figueroa},
  {Fujimoto}, {Fukazawa}, {Furusho}, {Furuzawa}, {Gendreau}, {Griffiths},
  {Haba}, {Hamaguchi}, {Harrus}, {Hasinger}, {Hatsukade}, {Hayashida}, {Henry},
  {Hiraga}, {Holt}, {Hornschemeier}, {Hughes}, {Hwang}, {Ishida}, {Ishisaki},
  {Isobe}, {Itoh}, {Iyomoto}, {Kahn}, {Kamae}, {Katagiri}, {Kataoka},
  {Katayama}, {Kawai}, {Kilbourne}, {Kinugasa}, {Kissel}, {Kitamoto}, {Kohama},
  {Kohmura}, {Kokubun}, {Kotani}, {Kotoku}, {Kubota}, {Madejski}, {Maeda},
  {Makino}, {Markowitz}, {Matsumoto}, {Matsumoto}, {Matsuoka}, {Matsushita},
  {McCammon}, {Mihara}, {Misaki}, {Miyata}, {Mizuno}, {Mori}, {Mori}, {Morii},
  {Moseley}, {Mukai}, {Murakami}, {Murakami}, {Mushotzky}, {Nagase}, {Namiki},
  {Negoro}, {Nakazawa}, {Nousek}, {Okajima}, {Ogasaka}, {Ohashi}, {Oshima},
  {Ota}, {Ozaki}, {Ozawa}, {Parmar}, {Pence}, {Porter}, {Reeves}, {Ricker},
  {Sakurai}, {Sanders}, {Senda}, {Serlemitsos}, {Shibata}, {Soong}, {Smith},
  {Suzuki}, {Szymkowiak}, {Takahashi}, {Tamagawa}, {Tamura}, {Tamura},
  {Tanaka}, {Tashiro}, {Tawara}, {Terada}, {Terashima}, {Tomida}, {Torii},
  {Tsuboi}, {Tsujimoto}, {Tsuru}, {Turner}, {Ueda}, {Ueno}, {Ueno}, {Uno},
  {Urata}, {Watanabe}, {Yamamoto}, {Yamaoka}, {Yamasaki}, {Yamashita},
  {Yamauchi}, {Yamauchi}, {Yaqoob}, {Yonetoku}, \& {Yoshida}}]{SUZAKU}
{Mitsuda}, K., {Bautz}, M., {Inoue}, H., {Kelley}, R.~L., {Koyama}, K.,
  {Kunieda}, H., {Makishima}, K., {Ogawara}, Y., {Petre}, R., {Takahashi}, T.,
  {Tsunemi}, H., {White}, N.~E., {Anabuki}, N., {Angelini}, L., {Arnaud}, K.,
  {Awaki}, H., {Bamba}, A., {Boyce}, K., {Brown}, G.~V., {Chan}, K.-W.,
  {Cottam}, J., {Dotani}, T., {Doty}, J., {Ebisawa}, K., {Ezoe}, Y., {Fabian},
  A.~C., {Figueroa}, E., {Fujimoto}, R., {Fukazawa}, Y., {Furusho}, T.,
  {Furuzawa}, A., {Gendreau}, K., {Griffiths}, R.~E., {Haba}, Y., {Hamaguchi},
  K., {Harrus}, I., {Hasinger}, G., {Hatsukade}, I., {Hayashida}, K., {Henry},
  P.~J., {Hiraga}, J.~S., {Holt}, S.~S., {Hornschemeier}, A., {Hughes}, J.~P.,
  {Hwang}, U., {Ishida}, M., {Ishisaki}, Y., {Isobe}, N., {Itoh}, M.,
  {Iyomoto}, N., {Kahn}, S.~M., {Kamae}, T., {Katagiri}, H., {Kataoka}, J.,
  {Katayama}, H., {Kawai}, N., {Kilbourne}, C., {Kinugasa}, K., {Kissel}, S.,
  {Kitamoto}, S., {Kohama}, M., {Kohmura}, T., {Kokubun}, M., {Kotani}, T.,
  {Kotoku}, J., {Kubota}, A., {Madejski}, G.~M., {Maeda}, Y., {Makino}, F.,
  {Markowitz}, A., {Matsumoto}, C., {Matsumoto}, H., {Matsuoka}, M.,
  {Matsushita}, K., {McCammon}, D., {Mihara}, T., {Misaki}, K., {Miyata}, E.,
  {Mizuno}, T., {Mori}, K., {Mori}, H., {Morii}, M., {Moseley}, H., {Mukai},
  K., {Murakami}, H., {Murakami}, T., {Mushotzky}, R., {Nagase}, F., {Namiki},
  M., {Negoro}, H., {Nakazawa}, K., {Nousek}, J.~A., {Okajima}, T., {Ogasaka},
  Y., {Ohashi}, T., {Oshima}, T., {Ota}, N., {Ozaki}, M., {Ozawa}, H.,
  {Parmar}, A.~N., {Pence}, W.~D., {Porter}, F.~S., {Reeves}, J.~N., {Ricker},
  G.~R., {Sakurai}, I., {Sanders}, W.~T., {Senda}, A., {Serlemitsos}, P.,
  {Shibata}, R., {Soong}, Y., {Smith}, R., {Suzuki}, M., {Szymkowiak}, A.~E.,
  {Takahashi}, H., {Tamagawa}, T., {Tamura}, K., {Tamura}, T., {Tanaka}, Y.,
  {Tashiro}, M., {Tawara}, Y., {Terada}, Y., {Terashima}, Y., {Tomida}, H.,
  {Torii}, K., {Tsuboi}, Y., {Tsujimoto}, M., {Tsuru}, T.~G., {Turner},
  M.~J.~L.~., {Ueda}, Y., {Ueno}, S., {Ueno}, M., {Uno}, S., {Urata}, Y.,
  {Watanabe}, S., {Yamamoto}, N., {Yamaoka}, K., {Yamasaki}, N.~Y.,
  {Yamashita}, K., {Yamauchi}, M., {Yamauchi}, S., {Yaqoob}, T., {Yonetoku},
  D., \& {Yoshida}, A. 2007, \pasj, 59, 1

\bibitem[{{Nandra} {et~al.}(1997){Nandra}, {George}, {Mushotzky}, {Turner}, \&
  {Yaqoob}}]{Nandra97}
{Nandra}, K., {George}, I.~M., {Mushotzky}, R.~F., {Turner}, T.~J., \&
  {Yaqoob}, T. 1997, \apj, 477, 602

\bibitem[{{Nandra} {et~al.}(2007){Nandra}, {O'Neill}, {George}, \&
  {Reeves}}]{pexmon}
{Nandra}, K., {O'Neill}, P.~M., {George}, I.~M., \& {Reeves}, J.~N. 2007,
  \mnras, 382, 194

\bibitem[{{Netzer} {et~al.}(2003){Netzer}, {Kaspi}, {Behar}, {Brandt},
  {Chelouche}, {George}, {Crenshaw}, {Gabel}, {Hamann}, {Kraemer}, {Kriss},
  {Nandra}, {Peterson}, {Shields}, \& {Turner}}]{Netzer03}
{Netzer}, H., {Kaspi}, S., {Behar}, E., {Brandt}, W.~N., {Chelouche}, D.,
  {George}, I.~M., {Crenshaw}, D.~M., {Gabel}, J.~R., {Hamann}, F.~W.,
  {Kraemer}, S.~B., {Kriss}, G.~A., {Nandra}, K., {Peterson}, B.~M., {Shields},
  J.~C., \& {Turner}, T.~J. 2003, \apj, 599, 933

\bibitem[{{Noda} {et~al.}(2011){Noda}, {Makishima}, {Yamada}, {Torii},
  {Sakurai}, \& {Nakazawa}}]{nado2011}
{Noda}, H., {Makishima}, K., {Yamada}, S., {Torii}, S., {Sakurai}, S., \&
  {Nakazawa}, K. 2011, ArXiv e-prints

\bibitem[{{Reynolds}(1997)}]{Reynolds1997R}
{Reynolds}, C.~S. 1997, \mnras, 286, 513

\bibitem[{{Ross} \& {Fabian}(2005)}]{reflionx}
{Ross}, R.~R. \& {Fabian}, A.~C. 2005, \mnras, 358, 211

\bibitem[{{Shu} {et~al.}(2010){Shu}, {Yaqoob}, \& {Wang}}]{Shu2010}
{Shu}, X.~W., {Yaqoob}, T., \& {Wang}, J.~X. 2010, \apjs, 187, 581

\bibitem[{{Takahashi} {et~al.}(2007){Takahashi}, {Abe}, {Endo}, {Endo}, {Ezoe},
  {Fukazawa}, {Hamaya}, {Hirakuri}, {Hong}, {Horii}, {Inoue}, {Isobe}, {Itoh},
  {Iyomoto}, {Kamae}, {Kasama}, {Kataoka}, {Kato}, {Kawaharada}, {Kawano},
  {Kawashima}, {Kawasoe}, {Kishishita}, {Kitaguchi}, {Kobayashi}, {Kokubun},
  {Kotoku}, {Kouda}, {Kubota}, {Kuroda}, {Madejski}, {Makishima}, {Masukawa},
  {Matsumoto}, {Mitani}, {Miyawaki}, {Mizuno}, {Mori}, {Mori}, {Murashima},
  {Murakami}, {Nakazawa}, {Niko}, {Nomachi}, {Okada}, {Ohno}, {Oonuki}, {Ota},
  {Ozawa}, {Sato}, {Shinoda}, {Sugiho}, {Suzuki}, {Taguchi}, {Takahashi},
  {Takahashi}, {Takeda}, {Tamura}, {Tamura}, {Tanaka}, {Tanihata}, {Tashiro},
  {Terada}, {Tominaga}, {Uchiyama}, {Watanabe}, {Yamaoka}, {Yanagida}, \&
  {Yonetoku}}]{SUZ_HXD}
{Takahashi}, T., {Abe}, K., {Endo}, M., {Endo}, Y., {Ezoe}, Y., {Fukazawa}, Y.,
  {Hamaya}, M., {Hirakuri}, S., {Hong}, S., {Horii}, M., {Inoue}, H., {Isobe},
  N., {Itoh}, T., {Iyomoto}, N., {Kamae}, T., {Kasama}, D., {Kataoka}, J.,
  {Kato}, H., {Kawaharada}, M., {Kawano}, N., {Kawashima}, K., {Kawasoe}, S.,
  {Kishishita}, T., {Kitaguchi}, T., {Kobayashi}, Y., {Kokubun}, M., {Kotoku},
  J., {Kouda}, M., {Kubota}, A., {Kuroda}, Y., {Madejski}, G., {Makishima}, K.,
  {Masukawa}, K., {Matsumoto}, Y., {Mitani}, T., {Miyawaki}, R., {Mizuno}, T.,
  {Mori}, K., {Mori}, M., {Murashima}, M., {Murakami}, T., {Nakazawa}, K.,
  {Niko}, H., {Nomachi}, M., {Okada}, Y., {Ohno}, M., {Oonuki}, K., {Ota}, N.,
  {Ozawa}, H., {Sato}, G., {Shinoda}, S., {Sugiho}, M., {Suzuki}, M.,
  {Taguchi}, K., {Takahashi}, H., {Takahashi}, I., {Takeda}, S., {Tamura},
  K.-I., {Tamura}, T., {Tanaka}, T., {Tanihata}, C., {Tashiro}, M., {Terada},
  Y., {Tominaga}, S., {Uchiyama}, Y., {Watanabe}, S., {Yamaoka}, K.,
  {Yanagida}, T., \& {Yonetoku}, D. 2007, \pasj, 59, 35

\bibitem[{{Taylor} {et~al.}(2003){Taylor}, {Uttley}, \& {McHardy}}]{Taylor03}
{Taylor}, R.~D., {Uttley}, P., \& {McHardy}, I.~M. 2003, \mnras, 342, L31

\bibitem[{{Terashima} {et~al.}(2008){Terashima}, {Gallo}, {Inoue}, {Markowitz},
  {Reeves}, {Anabuki}, {Fabian}, {Griffiths}, {Hayashida}, {Itoh}, {Kokubun},
  {Kubota}, {Miniutti}, {Takahashi}, {Yamauchi}, \& {Yonetoku}}]{Terashima08}
{Terashima}, Y., {Gallo}, L.~C., {Inoue}, H., {Markowitz}, A.~G., {Reeves},
  J.~N., {Anabuki}, N., {Fabian}, A.~C., {Griffiths}, R.~E., {Hayashida}, K.,
  {Itoh}, T., {Kokubun}, N., {Kubota}, A., {Miniutti}, G., {Takahashi}, T.,
  {Yamauchi}, M., \& {Yonetoku}, D. 2008, ArXiv e-prints

\bibitem[{{Uttley} {et~al.}(2004){Uttley}, {Taylor}, {McHardy}, {Page},
  {Mason}, {Lamer}, \& {Fruscione}}]{Uttley04}
{Uttley}, P., {Taylor}, R.~D., {McHardy}, I.~M., {Page}, M.~J., {Mason}, K.~O.,
  {Lamer}, G., \& {Fruscione}, A. 2004, \mnras, 347, 1345

\bibitem[{{Vaughan} \& {Fabian}(2004)}]{Vaug04}
{Vaughan}, S. \& {Fabian}, A.~C. 2004, \mnras, 348, 1415

\bibitem[{{Vaughan} {et~al.}(2003){Vaughan}, {Fabian}, \&
  {Nandra}}]{Vaughan_Fabian03}
{Vaughan}, S., {Fabian}, A.~C., \& {Nandra}, K. 2003, \mnras, 339, 1237

\bibitem[{{Vestergaard} \& {Peterson}(2006)}]{VestergaardPeterson2006}
{Vestergaard}, M. \& {Peterson}, B.~M. 2006, \apj, 641, 689

\bibitem[{{Walton} {et~al.}(2010){Walton}, {Reis}, \&
  {Fabian}}]{waltonreis2010}
{Walton}, D.~J., {Reis}, R.~C., \& {Fabian}, A.~C. 2010, \mnras, 408, 601

\bibitem[{{Zdziarski} {et~al.}(2003){Zdziarski}, {Lubi{\'n}ski}, {Gilfanov}, \&
  {Revnivtsev}}]{zdziarskietal2003}
{Zdziarski}, A.~A., {Lubi{\'n}ski}, P., {Gilfanov}, M., \& {Revnivtsev}, M.
  2003, \mnras, 342, 355

\bibitem[{{Zoghbi} {et~al.}(2010){Zoghbi}, {Fabian}, {Uttley}, {Miniutti},
  {Gallo}, {Reynolds}, {Miller}, \& {Ponti}}]{Zoghbi10}
{Zoghbi}, A., {Fabian}, A.~C., {Uttley}, P., {Miniutti}, G., {Gallo}, L.~C.,
  {Reynolds}, C.~S., {Miller}, J.~M., \& {Ponti}, G. 2010, \mnras, 401, 2419

\end{thebibliography}

\end{document}